\documentclass[pre,aps,floatfix,showpacs,onecolumn,preprint,superscriptaddress,12pt]{revtex4-1}
\pdfoutput=1
\usepackage{amssymb,amsfonts,amsmath}
\usepackage{graphics}
\usepackage[dvips]{graphicx}
\usepackage{color}
\usepackage{subfigure}
\usepackage{mathtools}
\usepackage[normalem]{ulem}
\usepackage{dcolumn}
\usepackage{multirow}
\usepackage{booktabs}
\usepackage{bm}
\usepackage{cancel}
\usepackage[colorinlistoftodos]{todonotes}
\usepackage{rotating}
\usepackage{pbox}    
\usepackage[linktocpage,colorlinks=true,linkcolor=blue,citecolor=blue]{hyperref}
\newcommand{\mf}[1]{\ensuremath{\mathbf{#1}}}

\newcommand{\mrm}[1]{\ensuremath{\mathrm{#1}}}

\newcommand{\unv}[1]{\ensuremath{\widehat{\mf{#1}}}}

\begin{document}

\title{Many-body microhydrodynamics of colloidal particles with active boundary layers}

% AUTHORS, AFFILIATIONS ================================================
\author{Rajesh Singh}
\affiliation{The Institute of Mathematical Sciences, CIT Campus, Chennai 600113, India}
\author{Somdeb Ghose}
\affiliation{The Institute of Mathematical Sciences, CIT Campus, Chennai 600113, India}
\author{R. Adhikari}
\affiliation{The Institute of Mathematical Sciences, CIT Campus, Chennai 600113, India}
\date{\today}
% ABSTRACT
% ========================================================================================
\begin{abstract}
Colloidal particles with active boundary layers - regions surrounding the
particles where non-equilibrium processes produce large velocity gradients -
are common in many physical, chemical and biological contexts. The velocity or
stress at the edge of the boundary layer determines the exterior fluid flow
and, hence, the many-body interparticle hydrodynamic interaction. Here, we
present a method to compute the many-body hydrodynamic interaction between $N$
spherical active particles induced by their exterior microhydrodynamic flow.
First, we use a boundary integral representation of the Stokes equation to
eliminate bulk fluid degrees of freedom. Then, we expand the boundary
velocities and tractions of the integral representation in an
infinite-dimensional basis of tensorial spherical harmonics and, on enforcing
boundary conditions in a weak sense on the surface of each particle, obtain a
system of linear algebraic equations for the unknown expansion coefficients.
The truncation of the infinite series, fixed by the degree of accuracy
required, yields a finite linear system that can be solved accurately and
efficiently by iterative methods. The solution linearly relates the unknown
rigid body motion to the known values of the expansion coefficients, motivating
the introduction of propulsion matrices. These matrices completely characterize
hydrodynamic interactions in active suspensions just as mobility matrices
completely characterize hydrodynamic interactions in passive suspensions. The
reduction in the dimensionality of the problem, from a three-dimensional
partial differential equation to a two-dimensional integral equation, allows
for dynamic simulations of hundreds of thousands of active particles on
multi-core computational architectures. In our simulation of $10^4$
active colloidal particle in a harmonic trap, we find that the necessary and 
sufficient ingredients to obtain steady-state convective currents, the so-called 
``self-assembled pump'', are (a) one-body self-propulsion and (b) two-body rotation from  
the vorticity of the Stokeslet induced in the trap. 
\end{abstract}


\maketitle

\tableofcontents
\section{Introduction}
% ============================================================================
There are many examples in physics, chemistry and biology, where non-equilibrium
processes at the surface of a particle drive exterior fluid flow and lead,
possibly, to motion of the particle. Such non-equilibrium processes are
frequently confined to a thin region surrounding the particle, as for example in
phoretic phenomena \cite{anderson1989colloid}, motion in chemically reacting
flows \cite{paxton2006chemical}, and in ciliary propulsion \cite{blake1971a}.
Following the pioneering efforts of Helmholtz \cite{helmholtz1879studien},
Smoluchowski \cite{smoluchowski1903} and Derjaguin \cite{derjaguin1947kinetic},
methods have been developed, to relate the rigid body motion of a particle to
the structure of the active boundary layer that surrounds it
\cite{anderson1989colloid}. The corresponding many-body problem, of determining
the collective motion and exterior flow of $N$ particles with active boundary
layers, has received much less attention \cite{anderson1981concentration,
chen1988electrophoresis, rider1993dynamic}.

In this paper, we present a systematic method of computing many-body
hydrodynamic interactions between colloidal particles due to the
exterior microhydrodynamic flow produced by active boundary layers at their
surfaces. The hydrodynamic interaction between particles is mediated by the
ambient fluid which, in the microhydrodynamic limit, obeys Stokes
equation. The boundary conditions are determined by the flow in the
active boundary layer. Depending on the structure of the boundary layer, these
may be Dirichlet conditions specifying the active surface velocity, or, Neumann
conditions specifying the surface traction. In either case, the fluid flow in
the bulk admits an integral representation in terms of the velocities and
tractions on the particle boundaries. We use the integral representation to
derive the rigid body motion of the particles, in terms of the boundary conditions
and any external forces and torques that may be applied to the particles.

In section \ref{section:boundary-integral} we present the Galerkin method of
solving the boundary integral equation. In this method, the boundary tractions
and velocities are expanded in an infinite-dimensional basis of complete,
orthogonal functions defined on the particle boundaries. 
On enforcing the boundary condition to this bulk fluid velocity, an integral
equation is obtained for the unknown boundary traction (when Dirichlet
conditions are specified) or the unknown boundary velocities (when Neumann
conditions are specified). The solution of this boundary integral equation
provides both the particle velocities and angular velocities and the bulk fluid
flow. On enforcing the boundary conditions in a weighted residual sense, with
weighting functions that are identical to the expansion functions, a system of
linear equations is obtained for the unknown expansion coefficients. The
solution of the linear system determines the expansion coefficients of the
unknown boundary traction (when Dirichlet conditions are specified) or that of
the unknown boundary velocities (when Neumann conditions are specified). We
focus attention on the former situation, where the velocity profile in the
boundary layer is provided. The velocity seen by the exterior fluid is the sum
of the rigid body motion of
the particle and the velocity at the outer edge of the boundary layer. When the
boundary layer is thin compared to the size of the particle, the boundary
condition can be applied directly on the surface of the particle as the sum of
a rigid body motion and an active slip. The problem is solved when the rigid
body motion and the exterior fluid flow are determined completely in terms of
the slip velocities specified on each particle surface. As the Stokes equations
are linear, the rigid body motion must be a linear functional of the active
slip. We show how this linear functional relationship is expressed through
propulsion matrices, which appear naturally in the solution of linear system of
equations. The propulsion matrices relate the vector of rigid body motions of
the particle to the vector of expansion coefficients of the active slip. The
problem of computing hydrodynamic interactions between active colloidal
particles is, thereby, reduced to that of computing the propulsion matrices.
The analysis also reveals that the propulsion
matrices are a sum of two parts : one is the contribution from the
superposition of flows due to each particle and the other is a correction
required to satisfy the boundary conditions. The first is the two-body
contribution to hydrodynamic interactions while the second contains the genuine
many-body contribution.

In section \ref{section:microhydrodynamics-active-spheres}, which contains the central 
results of this paper, we focus on spherical colloidal particles with active slip. We 
chose tensorial spherical harmonics as the Galerkin expansion basis, which are 
complete and orthogonal on the surface of the sphere. The simplicity of the spherical shape 
allows us to calculate all boundary integrals and matrix elements analytically, in terms of 
derivatives of the Green's function of Stokes flow. The structure of the flow has a pleasing
simplicity, as shown in Table \ref{table:irred_fluid_flow}. At any order, there are at most 
three kinds of derivatives, which are the irreducible gradient of the Green's function,
its curl and its Laplacian. This leads to a simple classification of the flow in terms of 
irreducible tensors of increasing rank. As numerical quadrature is no longer necessary
to evaluate the matrix elements, the linear system can be solved both efficiently and 
accurately. This linear solution yields the mobility and propulsion matrices which completely
describe the many-body hydrodynamic interactions in active suspensions. 

In section \ref{section:minimal-active-spheres} we truncate the infinite series expansion of 
the boundary fields to the minimal number of terms required to produce active
translations and rotations. At this order of truncation, all long-ranged 
hydrodynamic interactions are also automatically included. We show that there 
are exactly two coefficients in the expansion of the active slip that produce 
translation and rotation. The translational coefficient produces a flow that 
decays inverse cubically with distance while the
rotational coefficient produces a flow that decays inverse quartically with 
distance. We plot the flows generated by all terms in the truncated expansion 
in figure \ref{fig:single-particle-flows}. 

In section \ref{section:squirmers_trap} we illustrate our general formalism with 
a detailed study of the dynamics of active colloidal particles in a harmonic trap. For the squirming 
motion of a sphere, we recast the leading terms of Lighthill \cite{lighthill1952} and 
Blake's \cite{blake1971a} solution in our formalism. The problem of squirmers in a harmonic 
potential has been studied earlier by Nash et al. \cite{nash2010}, and more recently, by 
Hennes, Wolff and Stark \cite{hennes2014self}. In our study, we attempt to find out the 
necessary and sufficient ingredients to obtain steady-state convective currents, the 
so-called ``self-assembled pump''. We find that the key ingredients necessary for the
pumping state are (a) one-body self-propulsion and 
(b) two-body rotation from the vorticity of the Stokeslet induced 
in the trap. Thus, neither tumbling (as included in Nash et al.) nor stresslet 
flows (as included in Hennes et al.) are necessary. 
Our simulation of $10^4$ squirmers, the largest such simulation till date, displays the 
formation of the self-assembled pump with greater clarity than previous studies.

We conclude with a comparison of our approach with existing theories of collective 
hydrodynamics of active particles and 
with a discussion of applications of the integral equation technique to rheology in 
active colloidal suspensions.

%
% ============================================================================
\section{Boundary integral equation for microhydrodynamics}\label{section:boundary-integral}
% ============================================================================
We consider $N$ active particles, of arbitrary shape, in an incompressible fluid of viscosity $\eta$. The position of the center of mass, the translational velocity and the
rotational velocity about the center of the mass are $\mf R_n$, $\mf V_n$ and $\mf \Omega_n$ respectively. Points on the particle boundary $S_n$ are labelled, in the frame attached
to the center of the mass, by $\bm \rho_n$, or equivalently, in the laboratory frame by $\mf r_n = \mf R_n + \bm \rho_n$. 
Particle trajectories are obtained from Newton's equations, where the right hand sides include both
contact and body contributions,
\begin{align}
M\ \dot{\mf V}_n      = 
        \mf F_n+\mf F_n^{\mrm e}
,\qquad
I\ \dot{\mf \Omega}_n = 
        \mf T_n + \mf T_n^{\mrm e}.
\end{align}
The stress $\bm{\sigma}$ of the ambient fluid provides
the surface traction $\mf n\cdot\bm\sigma_n $ on each particle, from which the net contact force 
$\mf F_n =\int\mf n\cdot\bm\sigma_n \,\mrm dS_n$ and the net contact torque 
$ \mf T_n = \int\bm{\rho_n}\times (\mf n\cdot\bm\sigma_n)\, \mrm dS_n $ can be computed. 
Here $\mf n$ is the unit normal pointing away from the particle into the fluid,
$M$ and $I$ are the mass and the moment of inertia of the particle respectively,
and $\mf F^e$ and $\mf T^e$ are external body forces and torques. In the
absence of inertia, as appropriate to the microhydrodynamic regime, 
Newton's equations reduce to a pair of constraints, 
\begin{align}
        \mf F_n +\mf F_n^{\mrm e}= 0, \qquad
        \mf T_n+ \mf T_n^{\mrm e}= 0.
\label{eq:newtons_equations_constraints}
\end{align}
There is an instantaneous balance between contact and body forces at all times,
which precludes any acceleration of the particles. In the absence of external
forces, Newton's equations reduce further to  
\begin{align}
\mf F_n = 0, \qquad
\mf T_n = 0.
\label{eq:newtons_equations_force_torque_free_constraints}
\end{align}
The trivial solution to this is $\bm\sigma = 0$, that is, there is no motion in
the absence of external forces and torques. The non-trivial solutions describe
active motion, in which translation and rotation are possible in the absence of
external forces and torques. It is these non-trivial solutions that we seek
here. The velocity and angular velocity, having being determined from
non-trivial solutions of the stress, determine the evolution of the positions
and orientations through the kinematical equations, 
\begin{align} \dot{\mf{R}}_n = \mf{V}_n, \quad \dot{\mf{p}}_n = \bm{\Omega}_n
\times {\mf{p}}_n,
\end{align}
where ${\mf p}_n$ is a fixed axis passing through the center of mass of the
body. If the shape is not a figure of revolution about this axis, two
additional evolution equations, which together describe the motion of the orthogonal triad
attached to the center of mass, are necessary.

The contact forces on the particles are determined from the state of flow
in the ambient fluid. In the region bounded by the particles, the fluid satisfies
Stokes equation \cite{landau1987fluid, batchelor2000introduction},
\begin{subequations} \label{eq:stokes_equation} 
\begin{align} & \bm{\nabla}
\cdot \mf{v} = 0 , \quad\bm{\nabla} \cdot \bm{\sigma} = 0,\\
   % - \bm{\nabla} p + \eta \nabla^2 \mf{v} = 0
    %\\
   &\bm{\sigma}  =  -p \mathbb{I} + \eta \left( \bm{\nabla} \mf{v} +
   \bm{\nabla} \mf{v}^{\mrm T} \right),
\end{align}  \end{subequations}
where $\mf{v}$, $p$ and $\bm{\sigma}$ are the fluid velocity, pressure and
stress respectively. The solution of this equation, with the appropriate
boundary conditions, provides the surface traction and hence the contact
forces and torques that determine the rigid body motion.  

As discussed in the Introduction, a generic consequence of activity on the surface of rigid bodies
is the appearance of a boundary layer. When the boundary layer is thin
compared to the particle size, its effect appears as a boundary condition, on
either the velocity or the traction, at the particle surface. A detailed
discussion of both forms of boundary conditions and their applicability to
different boundary layer problems is available in \cite{anderson1989colloid}. 

The problem of determining the surface traction is substantially simplified by recognising
that the solution of the three-dimensional partial differential
equation in (\ref{eq:stokes_equation}) can be expressed as an integral of the velocity and traction
fields over the boundaries of the particles
\cite{fkg1930bandwertaufgaben, ladyzhenskaya1969, pozrikidis1992, kim2005}
\begin{align} \label{eq:many_body_boundary_integral_formulation}
  8 \pi \eta \, v_i(\mf{r})
  &=
  - \sum_{m=1}^N \int \left[
  G_{ij}(\mf{r},\mf{r}_m )f_j(\mf{r}_m) -  \eta \, K_{jik}(\mf{r},\mf{r}_m )n_k v_j(\mf{r}_m) \right] \, \mrm{d} S_m,
  \quad 
  \substack{
  \displaystyle \mf{r} \in V 
  \\ \\ 
  \displaystyle \quad\,\mf{r}_m \in S_m
  }
%   }
\end{align}
\begin{subequations}
\begin{align}
\nabla_j G_{ij}(\mf{r},\mf r' )  = 0, 
\end{align}
\begin{align}
-\nabla_i p_j(\mf{r},\mf r' )  + \nabla^2 G_{ij}(\mf{r},\mf r' )  = -\delta\left(\mf r-\mf r'\right)\delta_{ij},
\end{align}
\begin{align}
K_{ijk}(\mf{r},\mf r' ) = -\delta_{ik} \, p_j + \nabla_k G_{ij} 
                                              + \nabla_i G_{kj}.
\end{align}
\end{subequations}
Here $ \bm{\rho}=\mf{r} - \mf r'$ and $\mf f(\mf r_m) = \mf n\cdot\bm\sigma$
and $\mf v(\mf r_m)$ are, respectively, the traction and velocity on the boundary of the $m$-th
particle. $\mf G(\mf r, \mf r')$ is a
Green's function of Stokes flow, $\mf p$ is the
corresponding pressure vector and $\mf K(\mf r, \mf
r')$ is the stress tensor associated with this Green's function. The contribution from the boundary which
encloses both the fluid and the particles is not included here, since it is assumed that both the Green's
function and the flow vanish on this boundary. 

Equating the expression for the bulk fluid velocity to the boundary condition
and evaluating the second integral as a principal value leads to an integral
equation on the particle boundaries \cite{fkg1930bandwertaufgaben,
ladyzhenskaya1969, pozrikidis1992, kim2005},
\begin{align} \label{eq:many_body_boundary_integral_equation}
  4 \pi \eta \, v_i(\mf{r}_n)
  &=
  - \sum_{m=1}^N \int \left[
  G_{ij}(\mf{r}_n,\mf{r}^{\prime}_m )f_j(\mf{r}^{\prime}_m) -  
  \eta\, K_{jik}(\mf{r}_n,\mf{r}^{\prime}_m )n_k v_j(\mf{r}^{\prime}_m) \right] \, \mrm{d} S_m,
  \quad
  \substack{\displaystyle \mf{r}_n \in S_n \\ \\ \displaystyle \,\,\mf{r}^{\prime}_m \in S_m}.
\end{align}
For Dirichlet boundary conditions, this is a self-adjoint Fredholm integral
equation of the first kind for the unknown boundary tractions. For Neumann
boundary conditions, this is a self-adjoint Fredholm integral equation of the
second kind for the unknown boundary velocities. In both cases, the solution
linearly relates the surface traction to the surface velocity and the balance
of contact and body forces then determines the rigid body motion. The problem
of determining the traction is thus reduced to solving a two-dimensional
integral equation on the boundaries of the domain instead of a
three-dimensional partial differential equation in the bulk. 

Here we use the Galerkin method to discretize and solve the boundary integral
equation. In this method, the boundary fields are expanded in a complete,
orthogonal basis of functions $\phi^{(l)}(\bm \rho_n)$,
\begin{subequations}
\begin{align}
  \label{eq:f_galerkin_expansion}
  \mf{f}(\mf R_n + \bm\rho_n)
  &=
  \sum_{l = 0}^{\infty}
  \mf{F}_n^{(l)}
  {\phi}^{(l)}(\bm \rho_n),
\\
  \label{eq:v_galerkin_expansion}
  \mf{v}(\mf R_n + \bm\rho_n)
  &=
  \sum_{l = 0}^{\infty}
  \mf{V}_n^{(l)}
  {\phi}^{(l)}(\bm \rho_n).
\end{align}
\end{subequations}
The coefficients corresponding to the constant function and the antisymmetric
linear function are, respectively, the force and torque in the traction
expansion and the velocity and angular velocity in the velocity expansion.
Inserting this in the boundary integral representation,
(\ref{eq:many_body_boundary_integral_formulation}), provides an expression for
the bulk flow  
\begin{align} \label{eq:galerkin_representation_of_linear_VF_relation_yl}
  8 \pi \eta \, \mf{v}(\mf{r})
  &=
  - \sum_{m=1}^N
  \sum_{l = 0}^{\infty}\left[
  \bm{G}^{(l)}(\mf{r} ,\, \mf{R}_m)\cdot  \mf{F}_m^{(l)} -\eta
  \bm{K}^{(l)}(\mf{r} ,\, \mf{R}_m)\cdot  \mf{V}_m^{(l)}\right],
\end{align}
%
% %
in terms of the coefficients of each surface mode
$\phi^{(l)}(\bm \rho_m)$ of the velocity and traction fields and the two boundary
 integrals
\begin{subequations}
\label{eq:flow-elements}
\begin{align}
\bm{G}^{(l)}(\mf{r} ,\, \mf{R}_m) &= \int \mf{G}(\mf{r}, \mf{R}_m + \bm{\rho}_m)
  \,
  {\phi}^{(l)} (\bm{\rho}_m) \, \mrm{d} S_m, \label{eq:flow-matrix-element}
 \\
\bm{K}^{(l)}(\mf{r} ,\, \mf{R}_m) &=
\int \mf{K}(\mf{r}, \mf{R}_m +\bm{\rho}_m)\cdot\mf n \,
  \,
  {\phi}^{(l)} (\bm{\rho}_m) \, \mrm{d} S_m.
\end{align}
\end{subequations}
To determine the traction coefficients in terms of the velocity coefficients (or vice versa) 
we insert the velocity and traction expansions in the boundary integral
equation, (\ref{eq:many_body_boundary_integral_equation}), multiply by the
$l$-th basis function and integrate both sides over the $n$-th boundary.The
boundary conditions are thus enforced in a weighted integral, or weak, sense and
not point wise. The expansion and weighting basis functions are chosen to be
identical. This Galerkin procedure yields an infinite
dimensional system of linear equations which relate the velocity and traction coefficients

\begin{align} \label{eq:galerkin_representation_of_linear_VF_relation}
   4 \pi \eta \,
  \mf{V}_n^{(l)}&
  = 
  -
  \sum_{m=1}^N
  \sum_{l' = 0}^{\infty}\left[
  \bm{G}_{nm}^{(l, \, l')}(\mf{R}_n ,\, \mf{R}_m)
  \cdot\mf{F}_m^{(l')} - \eta  \bm{K}_{nm}^{(l, \, l')}(\mf{R}_n,\, \mf{R}_m)
  \cdot\mf{V}_m^{(l')}\right].
\end{align}
The matrix elements of this linear system are integrals of the Green's function
and the stress tensor over pairs of boundaries, weighted by the basis functions
on each boundary,
\begin{subequations}
\label{eq:matrix-elements}
\begin{align}
  \bm{G}_{nm}^{(l, \, l')}(\mf{R}_n ,\, \mf{R}_m)&
  =
  \int%_{S_n} 
  \int%_{S_m}
  {\phi}^{(l)} (\bm{\rho}_n)
  \,
  \mf{G}(\mf{R}_n + \bm{\rho}_n, \mf{R}_m +  \bm{\rho}^{\prime}_m)
  \,
  {\phi}^{(l')} (\bm{\rho}^{\prime}_m) 
  \, \mrm{d} S_m \mrm{d} S_n, 
\end{align}
\begin{align}
  \bm{K}_{nm}^{(l, \, l')}(\mf{R}_n ,\, \mf{R}_m)&
  =
  \int%_{S_n} 
  \int%_{S_m}
  {\phi}^{(l)} (\bm{\rho}_n)
  \,
  \mf{K}(\mf{R}_n + \bm{\rho}_n, \mf{R}_m +  \bm{\rho}^{\prime}_m)\cdot\mf n\,
  \,
  {\phi}^{(l')} (\bm{\rho}^{\prime}_m) 
  \, \mrm{d} S_m \mrm{d} S_n.
\end{align}
\end{subequations}
This linear system is valid for any shape of particle, any geometry of the
enclosing boundary and for both Neumann and Dirichlet boundary conditions.  

For concreteness, we shall focus on Dirichlet boundary conditions in the
remaining part of this paper. In this case, the velocity seen by the fluid at
the outer edge of the boundary layer is the sum of the rigid body motion of the
particle and the asymptotic value, $\mf v^a$, of the velocity in the boundary
layer,
\begin{subequations} 
\label{eq:boundary_conditions}  
  \begin{align} 
    \mf{v}({\mf r})= {\mf V}_n + \bm{\Omega}_n\times{\bm \rho}_n + {\mf
    v}^a({\bm\rho}_n) , &\quad {\mf r} = {\mf R}_n + {\bm\rho}_n \in
    S_n,
  \end{align}  
\begin{align}
   \int \mf v\cdot \mf n \,\mrm d S_n = 0.
\label{eq:no_flux}
\end{align}
\end{subequations}
Here, the active velocity $\mf v^a$ is assumed given while the rigid body
motion, $\mf V$ and $\mf \Omega$, is to be determined for all $N$ particles.
The only constraint we place on the active velocity is that it conserves mass,
as reflected in the integral condition of (\ref{eq:no_flux}). This makes our theory
completely general. In  specific cases, the active velocity may be determined
by other local fields like the temperature (in thermophoresis), ionic species
(in electrophoresis) and chemical species (in diffusiophoresis). Here we assume
that any non-fluid degree of freedom necessary to specify the active velocity
has already been determined, and thus, our work treats the purely hydrodynamic
aspect of the problem.

The structure of the linear system for the Dirichlet problem can be better
understood by clearly separating the known quantities from the unknown
quantities in (\ref{eq:galerkin_representation_of_linear_VF_relation}).  In the
velocity expansion, the velocity and angular velocity are unknowns, while all
remaining coefficients are fixed by the active velocity. In the traction
expansion, the contact force and contact torque are determined from Newton's
equations and are, therefore, known quantities, while all remaining coefficients
are unknowns. Therefore, we seek to obtain the unknown velocity and angular
velocity in terms of the active velocity, constrained by the balance of contact
and external forces as required by Newton's equation. 

To the above end, we group the expansion coefficients for all particles into
vectors and separate them into ``lower'' and ``higher'' vectors 
\begin{subequations}
\begin{align}
\mathsf V^{\mrm L}& =
\left(\mf V_1 -\mf V_1^a,\ldots,\mf V_N -\mf V_N^a,\,
\mf \Omega_1 - \mf \Omega_1^a,\ldots,\mf \Omega_N - \mf \Omega_N^a\right)^T ,\\
\mathsf V^{\mrm H}& =\left(\cdots, \mf V^{(l)}_1,\ldots,\mf V^{(l)}_N,
\ldots\right)^T ,
\qquad \not\in \mf V,\,\mf \Omega, \\
 \mathsf F^{\mrm L}& =\left(\mf F_1,\ldots, \mf F_N,\, \mf T_1,
\ldots, \mf T_N \right)^T ,\\
\mathsf F^{\mrm H}& =\left(\cdots, \mf F^{(l)}_1,\ldots,\mf F^{(l)}_N,
\ldots\right)^T, 
\qquad \not\in \mf F,\,\mf T.
\end{align}
\end{subequations}
The ``lower'' vectors contain the constant and linear antisymmetric
coefficients, while the ``higher'' vectors contain all the remaining
coefficients. The ``lower'' vectors in the velocity expansion contain
contributions from both the rigid body motion and the active velocity, while
the ``higher'' vectors contain contributions from the active velocity alone.
In these variables, the boundary integral equation takes the form of a matrix
equation, 
\begin{align} \label{eq:many_body_general_VL_VH_QL_QH_relation}
  4\pi\eta 
  \left(
  \begin{array}{c}
  \mathsf V^{\mrm{L}}  \\    \hline    \mathsf V^{\mrm{H}}
  \end{array}  
  \right)  =- 
  \left(
  \begin{array}{c | c}
    \mathsf{G}^{\mrm{LL}}     & \mathsf{G}^{\mrm{LH}}    \\    \hline    \mathsf{G}^{\mrm{HL}}     & \mathsf{G}^{\mrm{HH}}
  \end{array}
  \right)
  \left(
  \begin{array}{c}
    \mathsf F^{\mrm{L}}    \\    \hline    \mathsf F^{\mrm{H}}
  \end{array}
  \right)
+\eta
  \left(
  \begin{array}{c | c}
    \mathsf{K}^{\mrm{LL}}     & \mathsf{K}^{\mrm{LH}}    \\    \hline    \mathsf{K}^{\mrm{HL}}     & \mathsf{K}^{\mrm{HH}}
  \end{array}
  \right)
  \left(
  \begin{array}{c}
    \mathsf V^{\mrm{L}}    \\    \hline    \mathsf V^{\mrm{H}}
  \end{array}
  \right).
\end{align}
Here, the matrix elements that relate ``lower'' velocity vector to ``lower''
traction vector are collected together into $\mathsf G^{\mrm{LL}}$ and $\mathsf
K^{\mrm{LL}}$, and so on for the remaining three part of the linear system. The
linear system can be simplified by recalling that second term in
(\ref{eq:galerkin_representation_of_linear_VF_relation}) has the
eigenfunctions, $\mathsf K^{\mrm{LL}} \mathsf V^{\mrm L} = -4\pi \mathsf
V^{\mrm L}$, and produces no exterior flow for the rigid body component of the
motion, $\mathsf K^{\mrm {HL}} = 0$.  With these, the linear system becomes
\begin{subequations}
 \begin{align}
  8\pi\eta \,
  \mathsf V^{\mrm{L}}
  &=
  -\mathsf{G}^{\mrm{LL}}\, \mathsf F^{\mrm{L}}
  -\mathsf{G}^{\mrm{LH}}\, \mathsf F^{\mrm{H}}
  +\eta\mathsf{K}^{\mrm{LH}}\, \mathsf V^{\mrm{H}},
  \label{eq:many_body_active_passive_VL_FH_relation}
  \\
  4\pi\eta \,
  \mathsf V^{\mrm{H}}
  &=
  -\mathsf{G}^{\mrm{HL}} \, \mathsf F^{\mrm{L}}
  -\mathsf{G}^{\mrm{HH}} \, \mathsf F^{\mrm{H}}
  +\eta\mathsf{K}^{\mrm{HH}} \, \mathsf V^{\mrm{H}}.
  \label{eq:many_body_active_passive_VH_FH_relation}
\end{align} 
\end{subequations}
The unknown  traction coefficients are determined
from the solution of  (\ref{eq:many_body_active_passive_VH_FH_relation}), 
\begin{align}
\mathsf F^{\mrm{H}}
  &=
  - (\mathsf{G}^{\mrm{HH}})^{-1} 
  \big[\mathsf{G}^{\mrm{HL}}\, \mathsf F^{\mrm{L}} + 
  \left(4\pi\eta\, \mathbb{ I}
  -\eta\mathsf{K}^{\mrm{HH}} \right)
  \mathsf V^{\mrm{H}}\big].
  \label{eq:many_body_active_passive_stresses}
\end{align} 
Eliminating the unknown traction coefficients in
(\ref{eq:many_body_active_passive_VL_FH_relation}) gives the following expression
\begin{align}
  \Big[\mathsf{G}^{\mrm{LL}} -
  \mathsf{G}^{\mrm{LH}}\, 
   (\mathsf{G}^{\mrm{HH}})^{-1} 
  \mathsf{G}^{\mrm{HL}}\,\Big] \mathsf F^{\mrm{L}} 
  &=
-8\pi\eta \, \mathsf V^{\mrm{L}}
  +\Big[
 \eta\mathsf{K}^{\mrm{LH}}\, +
  \mathsf{G}^{\mrm{LH}}\, 
  (\mathsf{G}^{\mrm{HH}})^{-1} 
  \big(4\pi\eta\, \mathbb{ I}
  -\eta\mathsf{K}^{\mrm{HH}} \big)
  \Big] \mathsf V^{\mrm{H}}.
  \label{eq:many_body_active_passive_VL}
\end{align}
This formal solution achieves the objective of relating the known contact forces and torques, 
$\mathsf{F}^{\mrm L}$ to the rigid body motion, and the known
coefficients, $\mathsf{V}^{\mrm H}$, of the active velocity.  The solution can then be written as 
\begin{subequations}
\label{eq:many_body_active_FV}
\begin{align}
\sum_{m=1}^N \left[\bm\mu^{\mrm{TT}}_{nm}\cdot\mf F_m +
\bm\mu^{\mrm{TR}}_{nm}\cdot\mf T_m \right] 
  &= -\mf V_n     
+\sum_{m=1}^N\sum_{l=0}^{\infty} 
 \bm\pi^{(\mrm{T},\, l)}_{nm}\cdot \mf
V_m^{(l)},
\end{align}
\begin{align}
\sum_{m=1}^N \left[\bm\mu^{\mrm{RT}}_{nm}\cdot\mf F_m +
\bm\mu^{\mrm{RR}}_{nm}\cdot\mf T_m \right] 
  &=  -\mf \Omega_n
+\sum_{m=1}^N\sum_{l=0}^{\infty} 
\bm\pi^{(\mrm{R},\,l)}_{nm}\cdot \mf
V_m^{(l)}.
\label{}
\end{align}
\end{subequations}
Here, $\bm\mu$ are the mobility matrices familiar from the theory of passive
suspensions \cite{mazur1982, nunan1984effective, felderhof1977hydrodynamic,
cichocki1994friction, ichiki2002improvement, ladd1988, durlofsky1987dynamic,
brady1988dynamic}. The $\bm\pi$ are propulsion matrices, introduced here for the
first time, which relate the rigid body motion to modes of the active velocity.
Eliminating the contact forces and torques in favour of the body forces and torques using
(\ref{eq:newtons_equations_constraints}), the formal solution for the rigid body motion 
can be expressed as
\begin{subequations}
\label{eq:many_body_active_VL}
\begin{align}
\mf V_n      &= 
\sum_{m=1}^N \left[\bm\mu^{\mrm{TT}}_{nm}\cdot\mf F^e_m +
\bm\mu^{\mrm{TR}}_{nm}\cdot\mf T^e_m \right] 
+\sum_{m=1}^N\sum_{l=0}^{\infty} 
\bm\pi^{(\mrm{T},\, l)}_{nm}\cdot \mf
V_m^{(l)},
\end{align}
\begin{align}
\mf \Omega_n &= 
\sum_{m=1}^N \left[\bm\mu^{\mrm{RT}}_{nm}\cdot\mf F^e_m +
\bm\mu^{\mrm{RR}}_{nm}\cdot\mf T^e_m \right] 
+\sum_{m=1}^N\sum_{l=0}^{\infty} 
\bm\pi^{(\mrm{R},\,l)}_{nm}\cdot \mf
V_m^{(l)}.
\label{}
\end{align}
\end{subequations}
The above shows clearly that particles can both rotate and translate in the
absence of body forces and torques. In this force-free, torque-free scenario, it
is the propulsion matrices, and not the mobility matrices that determine the
hydrodynamic interaction of the particles. In contrast to the four mobility
matrices, there are, in principle, an infinite number of propulsion matrices,
corresponding to each distinct mode of the active velocity. Active hydrodynamic
interactions are, thus, intrinsically more rich and allow for a greater variety
in the dynamics than passive hydrodynamic interactions. The propulsion matrices
allow for a coupling between active translations and rotations which can lead to
non-rectilinear trajectories in the motion of even a single, isolated active
particle. 

The linear system is symmetric in both mode and particle indices, a property
that follows from the symmetry of the Green's function under interchange of
arguments and the tensorial indices. The positivity of energy dissipation ensures
that linear system has only positive eigenvalues. If the expansion is truncated
at $l=l_{max}$ the linear system has $N l_{max}$ unknowns and involves matrices
of size $Nl_{max}\times Nl_{max}$. 

The formal solution also shows that both the mobility and propulsion matrices
are always a sum of two parts : a direct superposition contribution
$(\mathsf{G}^{\mrm{LL}}$ and $\mathsf K^{\mrm{LH}})$ and many-body
contributions involving matrix inverses that ensure that the boundary
conditions are satisfied simultaneously on all particles. These many-body
contributions become increasingly important as the distance between particles
decreases. The formal solution provides a systematic method of obtaining the
many-body hydrodynamic interactions in colloidal particles with active
boundary layers, thereby, the problem is reduced to computing the mobility and
propulsion matrices. 

It is instructive to compare the Galerkin method employed here with the more
popular boundary element method of solution of Stokes flows. In Table
(\ref{tab:boundary-integtral-methods}) we have contrasted boundary integral
formulations and methods of discretization. Here, we have used a direct
formulation, in which the densities in
(\ref{eq:many_body_boundary_integral_equation}) are the physical quantities. In
indirect formulations, the boundary integral equation has either one of the
terms in (\ref{eq:many_body_boundary_integral_equation}), but not both. In
such formulations, the densities are not directly related to physical
quantities. The advantage in using indirect formulations is that they yield
better conditioned linear systems when discretized. The discretization of the
boundary integral equation, in either formulation, can be done either through a
collocation method, which enforces the equation point-wise in a strong sense,
or through a Galerkin method which enforces the equation as a weighted integral
in a weak sense. Youngren and Acrivos \cite{youngren1975stokes} used a direct
formulation of the boundary integral equation but used a collocation method of
solution. Zick and Homsy \cite{zick1982stokes} were the first to use a
Galerkin discretization but used an indirect formulation. Our work is, to the
best of our knowledge, the first Galerkin discretization using a direct
formulation. The direct formulation is essential with slip velocities and when
a formulation using physical quantities is desired. Galerkin methods yield
symmetric discretization of self-adjoint problems and are thus preferred, when
feasible, over collocation discretization which usually do not preserve
self-adjointness \cite{youngren1975stokes, power1987second}. For certain smooth
boundaries, for example spheres, Galerkin methods provides the most accurate
results for the least number of unknowns \cite{muldowney1995spectral}.

\begin{table}
\begin{center}
\begin{tabular}{|c|c|c|}
\hline
\multirow{2}{3cm}{\centering{\mf{ Discretization}}}& \multicolumn{2}{p{10cm}|}{
    \centering{\hspace{-0.8cm}\mf{~ ~ ~ ~ ~\qquad Formulation}}} \\\cline{2-3}
 & \multicolumn{1}{c|}{Direct (single {and} double layer) } &
 \multicolumn{1}{c|}{Indirect (single {or} double layer)} \\ \hline
 Collocation & Youngren and Acrivos (1975) & Power and Miranda (1987)    \\
 \hline
 Galerkin    & This {work} & Zick and Homsy (1982)   \\
\hline
\end{tabular}
\end{center}
\caption{ Methods for the solution of the boundary integral equation of the Stokes
    equation. In this work, we have done a direct, involving both single and the
    double layer, formulation of the boundary integral method with  Galerkin
    discretization.
    }
\label{tab:boundary-integtral-methods}
\end{table}
To summarize, the main result of this section is an explicit expression for the
rigid body motion of $N$ particles, in terms of the mobility and the propulsion
matrices. The propulsion matrices are infinite in number as compared to only
four mobility matrices and this explains the much richer dynamics and
interesting orbits seen in active particles, even in the absence of external
forces and torques. The mobility and propulsion matrices are obtained in terms
of matrix elements of the Green's function and the stress tensor, in a basis of
complete orthogonal functions on the particle boundaries. The results in this
section are valid, for any shape of particle and any geometry of the boundary
enclosing the fluid. In the next section, we specialize to spherical particles
and an unbounded fluid. For spherical particles, the boundary integrals and the
matrix elements can be expressed in terms of \emph{derivatives} of the Green's
function, and additionally, the one-particle matrix element is diagonal in an
unbounded fluid. The explicit analytical form of the matrix elements leads to
an efficient numerical method of simulating active suspensions, as we explain
below.  
%
% ============================================================================
\section{Microhydrodynamics of active spheres in an unbounded fluid}
\label{section:microhydrodynamics-active-spheres}
\begin{figure}
\begin{tikzpicture}[scale=0.9]
\node (s1) at (4, 6) {};   % sphere #1
\node (s2) at (10, 6) {};   % sphere #2

% Draw axes
\draw[very thick,->] (0,0,0) -- (0, 0, 12) node[anchor=east]     {$x$};
\draw[very thick,->] (0,0,0) -- (12, 0, 0) node[anchor=north east]{$ y$};
\draw[very thick,->] (0,0,0) -- (0, 7, 0) node[anchor=north east]{$z$};
\shade [ball color=black!25!] (s1) circle [radius=1cm];         % sphere #1
\shade [ball color=black!25!] (s2) circle [radius=1cm];         % sphere #2

\draw [very thick,->, color=black!90] (0, 0)--(3, 6) node [midway, above, sloped] {\textcolor{black}{$\mf r_n$}};
\draw [very thick,->, color=black!90] (0, 0)--(4, 6) node [midway, below, sloped] {\textcolor{black}{$\mf R_n$}};
\draw [very thick,->, color=black!66] (4, 6)--(3, 6) node [midway, above,
sloped] {\textcolor{black}{$\bm \rho_n$}};
\draw[->, >=latex, red!36!black!20, line width=3pt] (4, 6)--(4.5, 5.144) node[midway, above, black, sloped]{$\mf p_n$} ;
\draw [very thick,->, color=black!90] (0, 0)--(10, 5) node [midway, below, sloped] {\textcolor{black}{$\mf r_m$}};           
\draw [very thick,->, color=black!90] (0, 0)--(10, 6) node [midway, above, sloped] {\textcolor{black}{$\mf R_m$}};
\draw [very thick,->, color=black!66] (10, 6)--(10, 5) node[midway, above,
sloped] {\textcolor{black}{$\bm \rho_m$}};
\draw[->, >=latex, red!36!black!20, line width=3pt] (10, 6)--(10.707, 6.7077) node[midway, above, black, sloped]{$\mf p_m$} ;
\end{tikzpicture}
\caption{
    Coordinate system used to describe active spherical particles. The $m$-th
    and $n$-th particles are shown. Center of mass coordinates are ${\bf R}_m$
    and ${\bf R}_n$ while orientations are the unit vectors ${\bf p}_m$ and
    ${\bf p}_n$.  Points on the boundaries of the spheres are ${\bf r}_m = {\bf
    R}_m + \bm{\rho}_m$  and ${\bf r}_n = {\bf R}_n + \bm{\rho}_n$.
    } 
\label{coordinate-system}
\end{figure}
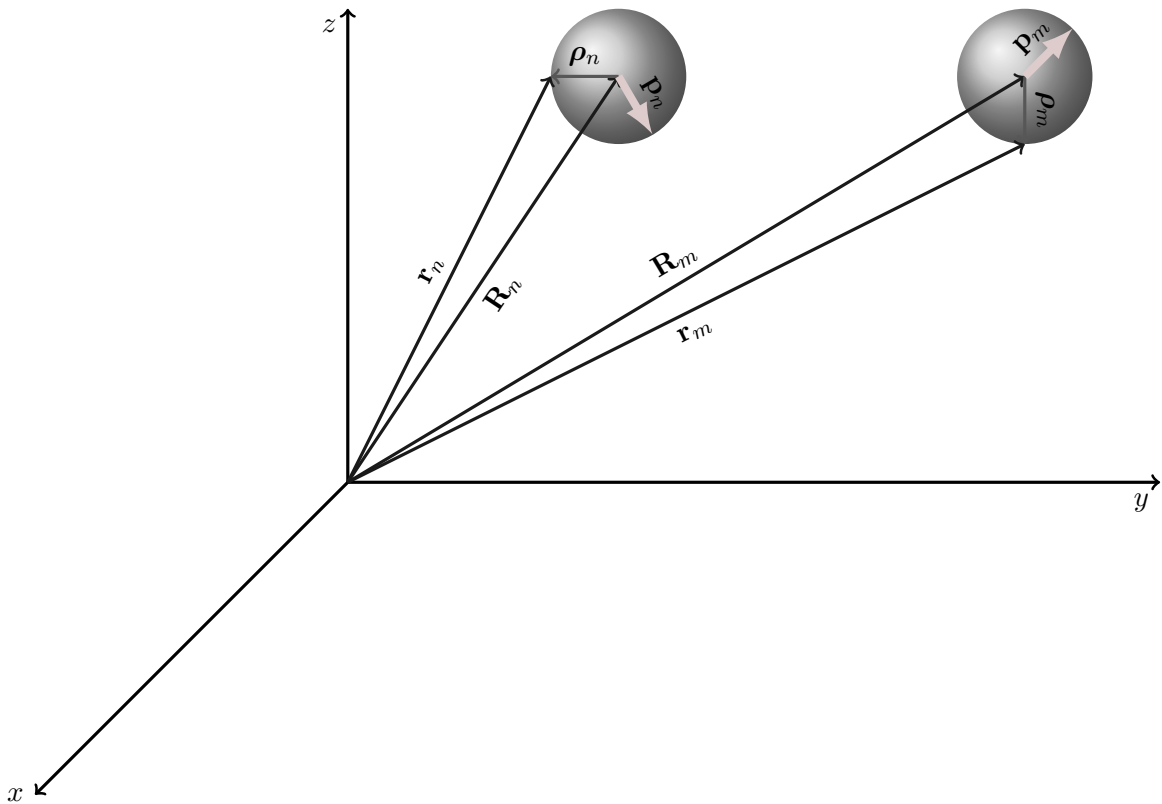

% ============================================================================
We now consider the problem of $N$ colloidal spheres of radius $a$ with active
boundary layers in an unbounded Stokes flow. The sphere centers are at $\mf
R_n$ and their velocities and angular velocities are $\mf V_n$ and $\mf
\Omega_n$ respectively. Additionally, the orientation of each sphere is
specified by an unit vector $\mf p_n$, which represents the symmetry axis of
the active velocity. The system of coordinates is shown in
figure~\ref{coordinate-system}. We assume Dirichlet boundary conditions on the
surface of each sphere and fluid to be at rest at infinity, 
\begin{subequations} 
\label{eq:boundary_conditions}  
  \begin{align} 
    \mf{v}({\mf r})= {\mf V}_n + \bm{\Omega}_n\times{\bm \rho}_n + {\mf
    v}^a({\bm\rho}_n) , &\quad {\mf r} = {\mf R}_n + {\bm\rho}_n \in
    S_n,
    \\
   \int \mf v\cdot \mf n \,\mrm d S_n = 0,
  \end{align}  
\end{subequations}
\begin{align}
    |{\mf v}| \rightarrow 0,\,\, |p|\rightarrow 0, &\quad |{\mf r}| \rightarrow \infty.
\label{}
\end{align}
We now obtain explicit expressions for the boundary integrals and matrix elements
from which the solution of (\ref{eq:stokes_equation}) can be obtained. 
The Green's function $\mf G(\mf r, \mf r')$ of Stokes 
flow that vanishes at infinity, $\mf p$ the
corresponding pressure vector, and $\mf K(\mf r, \mf
r')$ the stress tensor associated with this Green's function are
\begin{subequations}
\begin{align}
p_j(\mf{r}, \mf r' ) = -\nabla_j\nabla^2{\rho} = 2\frac{\rho_j}{\rho^3},
\end{align}
\begin{align}
G_{ij}(\mf{r}, \mf r' ) =\left(\delta_{ij}\nabla^2 -\nabla_i\nabla_j\right)
\rho = \frac{\delta_{ij}}{\rho} + \frac{\rho_i\rho_j}{\rho^3}, 
\end{align}
\begin{align}
K_{ijk}(\mf{r},\mf r' ) = -\delta_{ik} \, p_j + \nabla_k G_{ij} + \nabla_i
G_{jk} = -6\frac{\rho_i\rho_j\rho_k}{\rho^5},
\end{align}
\end{subequations}
where, as before, $ \bm{\rho}=\mf{r} - \mf r'$.
%
%
%%
%% ============================================================================
%% FIGURE # : COORDINATE SYSTEM
%% ============================================================================
%%
%
The natural choice of Galerkin basis functions $\phi^{(l)}(\bm \rho_n)$ on the
sphere are spherical harmonics. However, they are less convenient for
expanding vector fields as the expansion coefficients no longer transform as
Cartesian tensors under rotations. This inconvenience can be circumvented by
choosing tensorial spherical harmonics as the Galerkin expansion basis
\cite{weinert1980spherical, brunn1976effect, ghose2014irreducible}. The $l$-th
tensorial spherical harmonic is an irreducible Cartesian tensor of rank $l$.
Consequently, the expansion coefficients are Cartesian tensors of rank ($l+1)$,
symmetric irreducible in their last $l$ indices. Elementary angular momentum
algebra shows that they must each be a sum of three irreducible tensors of rank
$(l-1)$, $l$ and $(l+1)$ \cite{tung1985group}. This leads to a very convenient
classification of boundary integrals, matrix elements, and rigid body motion, as
we shall show below. 

The tensorial spherical harmonics are defined as 
\begin{align}\label{eq:tensorial_spherical_harmonics_definition} 
Y_{\alpha_1\alpha_2\dots\alpha_{l}}^{(l)}(\unv{\bm \rho}) &= (-1)^l \,
\rho^{l+1} \left( \nabla_{\alpha_1} \nabla_{\alpha_2} \dots
\nabla_{\alpha_l}\right) 
\frac{1}{\rho}.
\end{align}
We write this in a more compact notation
\begin{align}
\mf Y^{(l)}(\unv{\bm\rho}) &= (-1)^l \,\rho^{l+1} \bm\nabla^{(l)}
\frac{1}{\rho}.
\label{}
\end{align}
The tensorial spherical harmonics are orthogonal on the sphere \cite{hess1980formeln}, 
\begin{align}
\label{eq:tensorial_spherical_harmonics_orthogonality}
\frac{1}{4\pi a^2}\int\mf Y^{(l)}(\unv{\bm\rho}_n) \,\mf Y^{(l')}(\unv{\bm\rho}_n) \, \mrm dS & =
\delta_{ll'} \, \frac{l!\,(2l-1)!!}{(2l+1)}\mf \Delta^{(l)},
\end{align}
where $\mf \Delta ^{(l)}$ is a tensor of rank $2l$ that projects any $l$-th
order tensor to its symmetric irreducible form \cite{hess1980formeln,
mazur1982}. 

We expand both velocities and tractions on the surface of the particles in
tensorial spherical harmonics as
\begin{subequations}
\label{eq:VF_Yl_expansion}
\begin{align}
\mf f(\mf R_n +\bm\rho_n) &= \sum_{l=0}^{\infty} \frac{(2l+1)}{4\pi a^2}\,  
\mf F_n^{(l+1)}\odot\mf Y^{(l)}(\unv{\bm\rho}_n),
\label{eq:stress-spherical-harmonics-expansion}
\\
    {\mf v}(\mf R_n +\bm\rho_n) 
    &=  \, 
    \sum_{l = 0 }^{\infty}
    \frac{1}{ l!\, (2l -1)!!}\,
  {\mf V}_n^{(l+1)}  \odot  {\mf Y}^{(l)}(\unv{\bm\rho}_n),
    \label{eq:velocity-expansion-spherical-harmonics}
\end{align}
\end{subequations}
where $\odot$ indicates a $l$-fold contraction between a tensor of rank-$l$ and
a higher rank tensor. From the orthogonality of the tensorial harmonics, it follows that
\begin{subequations}\label{eq:expansion_coefficients}
\begin{align}
\mf F_n^{(l+1)} = \frac{1}{l!\,(2l-1)!!} \int \mf f(\mf R_n +\bm\rho_n) \,\mf
Y^{(l)}(\unv{\bm\rho}_n)\, \mrm d S_n,
\label{}
\end{align}
\begin{align}
    {\mf V}_n^{(l + 1)} 
    = & \,
    \frac{(2l +1)}{4 \pi a^{2} }
    \int_{S'} 
    {\mf v} (\mf R_n + \bm\rho_n)\, {\mf Y}^{(l)}(\unv{\bm\rho}_n)\, \mrm dS_n.
\label{}
\end{align}
\end{subequations}
From these definitions, it is clear that $\mf F_n^{(l+1)}$ and $\mf
V_n^{(l+1)}$ are tensors of rank $l+1$, symmetric irreducible in their last $l$
indices. As mentioned before, it follows from angular momentum algebra that $\mf
F^{(l+1)}$ and $\mf V^{(l+1)}$ can be expressed as the sum of three irreducible
tensors of rank $(l+1)$, $l$, and $(l-1)$. These are \cite{schmitz1980,
brunn1976effect}
\begin{subequations}
\begin{align}\label{eq:irred_traction_coefficients}
\mf F^{(l)} &=
\bm\Delta^{(l)}\mf F^{(l0)}
-\frac{l-1}{l}\bm\Delta^{(l-1)}        \bm\epsilon\cdot\mf F^{(l1)} +
\frac{l(l-1)}{2(2l-1)}\bm\Delta^{(l-1)}\bm\delta\,\mf F^{(l2)},
\end{align}
\begin{align}\label{eq:irred_velocity_coefficients}
\mf V^{(l)} &=
\bm\Delta^{(l)}                                    \mf V^{(l0)}
-\frac{l-1}{l}\bm\Delta^{(l-1)}        \bm\epsilon\cdot\mf V^{(l1)} +
\frac{l(l-1)}{2(2l-1)}\bm\Delta^{(l-1)}\bm\delta\,    \mf V^{(l2)},
\end{align}
\end{subequations}
where $\mf F^{(l\sigma)}$ and $\mf V^{(l\sigma)}$ are symmetric irreducible
tensors of rank $l-\sigma$. The irreducible parts are obtained from the
reducible tensors by complete symmetrization and detracing and by appropriate
contractions with the Levi-Civita and identity tensors, 
\begin{subequations}
\begin{align}\label{eq:irred_traction_coefficients_inverse}
\mf F^{(l0)} = \overbracket[0.7pt][2.0pt]{\mf F^{(l)}},\qquad
\mf F^{(l1)} = \overbracket[0.7pt][2.0pt]{\bm\epsilon\cdot\mf F^{(l)}},\qquad
\mf F^{(l2)} = {\bm\delta:\mf  F^{(l)}},
\end{align}
\begin{align}\label{eq:irred_velocity_coefficients_inverse}
\mf V^{(l0)} = \overbracket[0.7pt][2.0pt]{\mf V^{(l)}},\qquad
\mf V^{(l1)} = \overbracket[0.7pt][2.0pt]{\bm\epsilon\cdot\mf V^{(l)}},\qquad
\mf V^{(l2)} = {\bm\delta:\mf  V^{(l)}}.
\end{align}
\end{subequations}
Inserting the velocity and traction expansions in the boundary integral representation
gives the flow as
\begin{align} \label{eq:galerkin_representation_of_boundary_integral_formulation}
  8 \pi \eta \, \mf{v}(\mf{r})
  &=
  - \sum_{m=1}^N
  \sum_{l = 0}^{\infty}\left[
  \bm{G}^{(l+1)}(\mf{r} ,\, \mf{R}_m)\odot  \mf{F}_m^{(l+1)} -\eta
  \bm{K}^{(l+1)}(\mf{r} ,\, \mf{R}_m)\odot  \mf{V}_m^{(l+1)}\right],
\end{align}
where the boundary integrals are  
\begin{subequations}
\label{eq:flow-elements}
\begin{align}
\bm{G}^{(l+1)}(\mf{r} ,\, \mf{R}_m) &= 
    \frac{2l+1}{4\pi a^2}\,  
    \int \mf{G}(\mf{r}, \mf{R}_m + \bm{\rho}_m)  \,
    {\mf Y}^{(l)} (\unv{\bm\rho}) \, \mrm{d} S,
 \\
\bm{K}^{(l+1)}(\mf{r} ,\, \mf{R}_m) &=
    \frac{1}{l!\left(2l-1\right)!!}\,  
    \int \mf{K}(\mf{r}, \mf{R}_m + \bm{\rho}_m)\cdot \mf n  \,
    {\mf Y}^{(l)} (\bm{\rho}) \, \mrm{d} S. 
\end{align}
\end{subequations}
As we show in the Appendix \ref{appendix:flow_elements}, these boundary integrals can
be expressed as \emph{derivatives} of the Green's function 
and the stress tensor, 
\begin{subequations}\label{eq:flow_expression}
\begin{align}
\bm{G}^{(l+1)}(\mf{r} ,\, \mf{R}_m) =& 
    a^{l}\bm\Delta^{(l)}
    \left( 1 + \frac{a^2}{4l+6}    \nabla_m^2\right)
    \bm\nabla_m^{(l)}
    \mf G(\mf r, \,\mf R_m),
\end{align}
\begin{align}
    \bm K^{(l+1)}(\mf r,\,\mf R_m)
    =&
    \frac{4 \pi a^{l+1}\bm\Delta^{(l)} }{(l-1)!(2l +1)!!}
    \left( 1 + \frac{a^2}{4l+6}    \nabla_m^2\right)
    {\bm \nabla_m}^{(l-1)}
    \mf K(\mf r, \,\mf R_m),
\end{align}
\end{subequations}
where $\bm K^{(l+1)}(\mf r,\,\mf R_m) $ is defined for $l\geq 1$ and vanishes
identically for $l=0$. Using (\ref{eq:irred_traction_coefficients}) and
(\ref{eq:irred_velocity_coefficients}), the $l$-th order gradient in the first
flow integral can be decomposed into a sum of three irreducible gradients,
corresponding to each irreducible component of the coefficients. Further, as we
show in Appendix \ref{appendix:solution_green_function}, expressing the stress
tensor in terms of the Green's function and pressure vector, and utilizing the
equation of motion, (\ref{eq:newtons_equations_constraints}), that connects them,
the $l $-th term in the second flow integral can be expressed as gradients of
the Green's function alone. The irreducible decomposition of the gradient again
yields exactly three irreducible terms, corresponding to the irreducible
components of the velocity coefficients. Grouping both the first and second
series together, then provides a compact expression for the exterior flow due to
$N$ active spheres, 
\begin{align}
\label{eq:galerkin_representation_flow_yl}
\mf v(\mf r) = -\sum_{m}\sum_{l}\sum_{\sigma} \mf v^{(l\sigma)}(\mf r, \mf R_m),
\end{align}
\begin{align}
8 \pi \eta \, \mf v^{(l\sigma)}(\mf r, \mf r_m) = 
    \left\{ 
  \begin{array}{l l l }
    a^l\bm\Delta^{(l+1)}
    \mathcal{F}_m^l
    \bm\nabla_m^{(l)} \mf G(\mf r, \mf r_m)   \odot   \mf{Q}_m^{(l0)},
    \quad &\sigma=0
    \\\\
   -a^l\frac{l}{l+1}  \bm\Delta^{(l)}\left(\bm\nabla_m^{(l)} \times\mf G(\mf r, \mf r_m) \right)
   \odot   \mf{Q}_m^{(l1)},
    \quad &\sigma=1
   \\\\
   a^l\frac{l(l+1)}{2(2l+1)}
    \bm\Delta^{(l)}\bm\nabla_m^{(l-2)}\nabla_m^2 \mf G(\mf r, \mf r_m) 
    \odot   \mf{Q}_m^{(l2)},
    \quad &\sigma=2,
    \end{array} \right.
\end{align}
where 
\begin{align}
\mathcal{F}_n^l = \left(1 + \frac{a^2}{4l+6} \nabla_n^2\right),
\label{}
\end{align}
is an operator which encodes the finite size of the sphere. The flow is
expressed as irreducible gradients of the Green's function with coefficients
which are linear combinations of the irreducible traction and velocity
coefficients.  In this form, the flow field is manifestly incompressible and
biharmonic. Thus far, no use has been made of the properties of the Green's
function in an unbounded fluid, and thus, these results are valid for any
bounding geometry. The irreducible combinations of the velocity and traction
coefficients are 
\begin{align}
\mf Q^{(l\sigma)} =
    \left\{ 
  \begin{array}{l l l }
\mf{F}^{(l0)} - \left(1-\delta_{l0}\right)\frac{8 \pi\eta a}{(l-2)!(2l-1)!!}\mf{V}^{(l0)},
\qquad&\sigma=0
  \\\\
\mf{F}^{(l1)} - \frac{8(l-1)\pi\eta a}{l(l-2)!(2l-1)!!}\mf{V}^{(l1)},
\qquad&\sigma=1
\\\\
\mf{F}^{(l2)} + \frac{12\pi\eta a}{(l-2)!(2l-1)!!}\mf{V}^{(l2)},    
\qquad&\sigma=2.
\label{}
    \end{array} \right.
\label{eq:generalized_Qlsigma}
\end{align}
The contributions from the velocity vanish for $l=0$ and $l=1, \sigma = 1$, as
the second integral vanishes for a rigid body motion. For $l=1, \sigma = 0$, we
recognize the combination of symmetric traction and velocity moments first
introduced by Landau and Lifshitz \cite{landau1987fluid} and subsequently called
the stresslet by Batchelor \cite{batchelor1970}. The above provides a
generalization of this coefficient to arbitrary orders in tensorial harmonic
expansion. The flow due to the $l$-th term has contributions that decay as
$1/\rho^{l+1}$ and $1/\rho^{l+3}$. Thus, an expansion truncated at $l=2$ includes all
long-ranged contributions to the flow. 

\begin{figure}
\begin{center}
\includegraphics[width=0.496\textwidth]{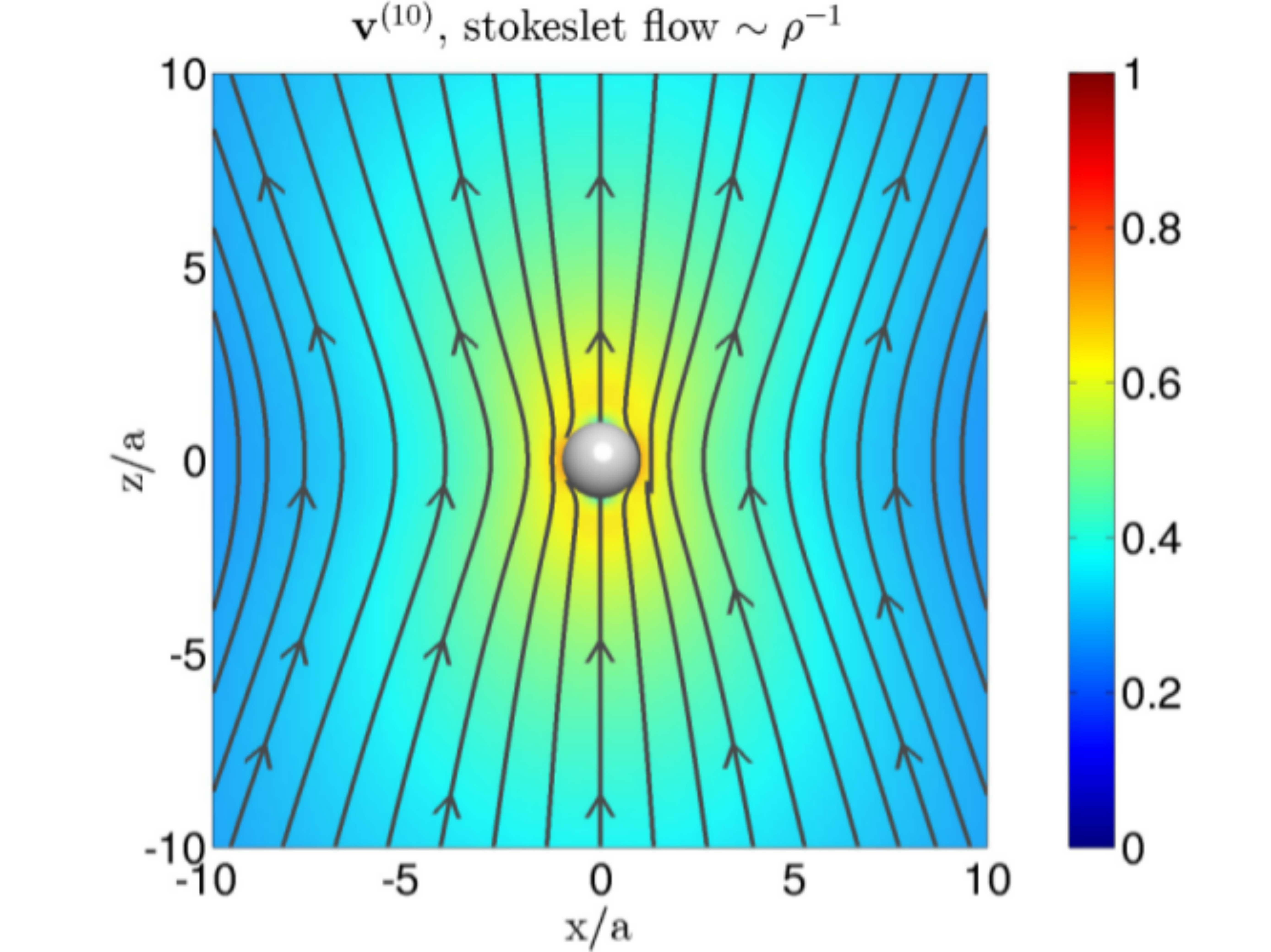} 
\includegraphics[width=0.496\textwidth]{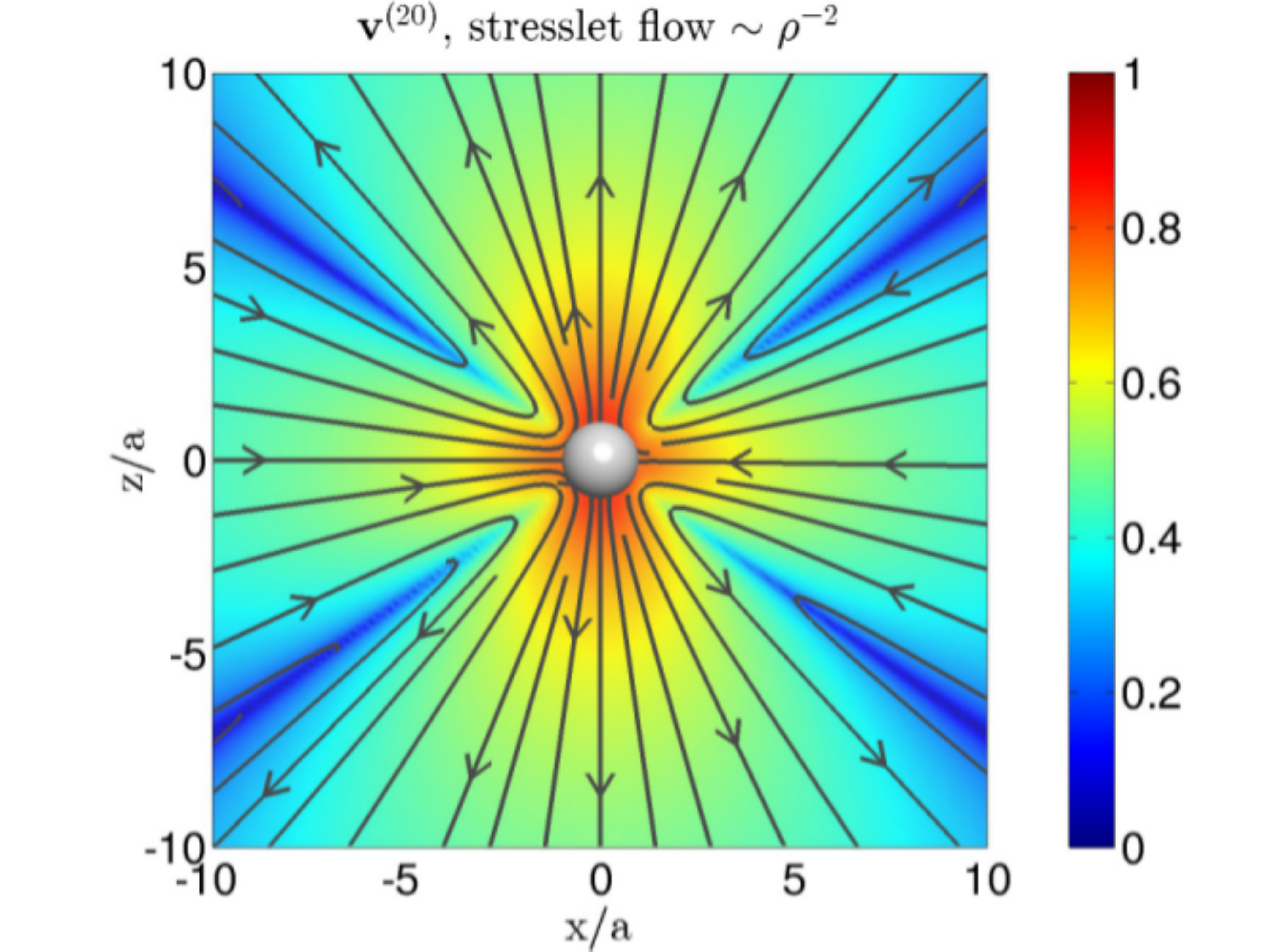}\\
\includegraphics[width=0.496\textwidth]{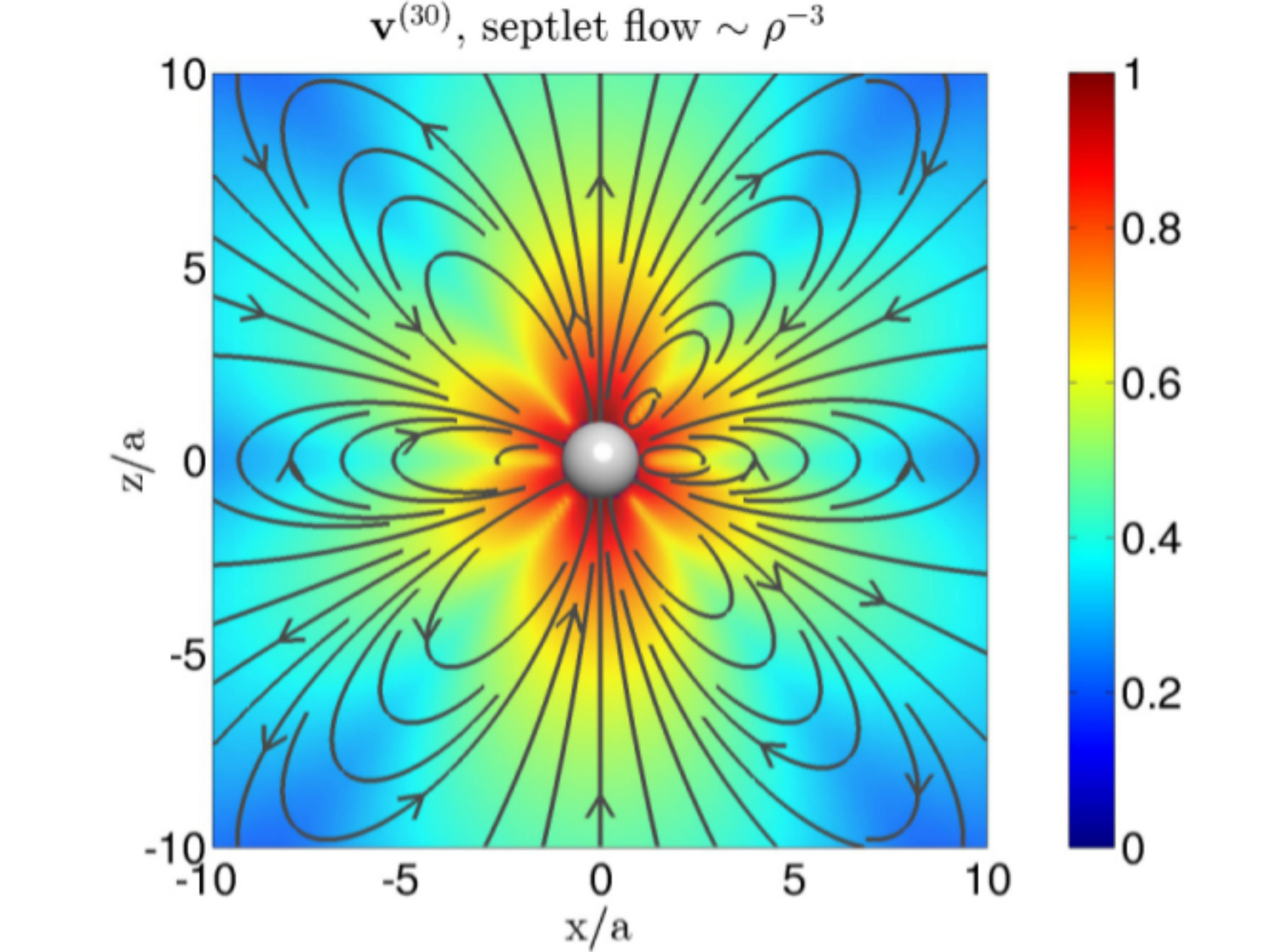}
\includegraphics[width=0.496\textwidth]{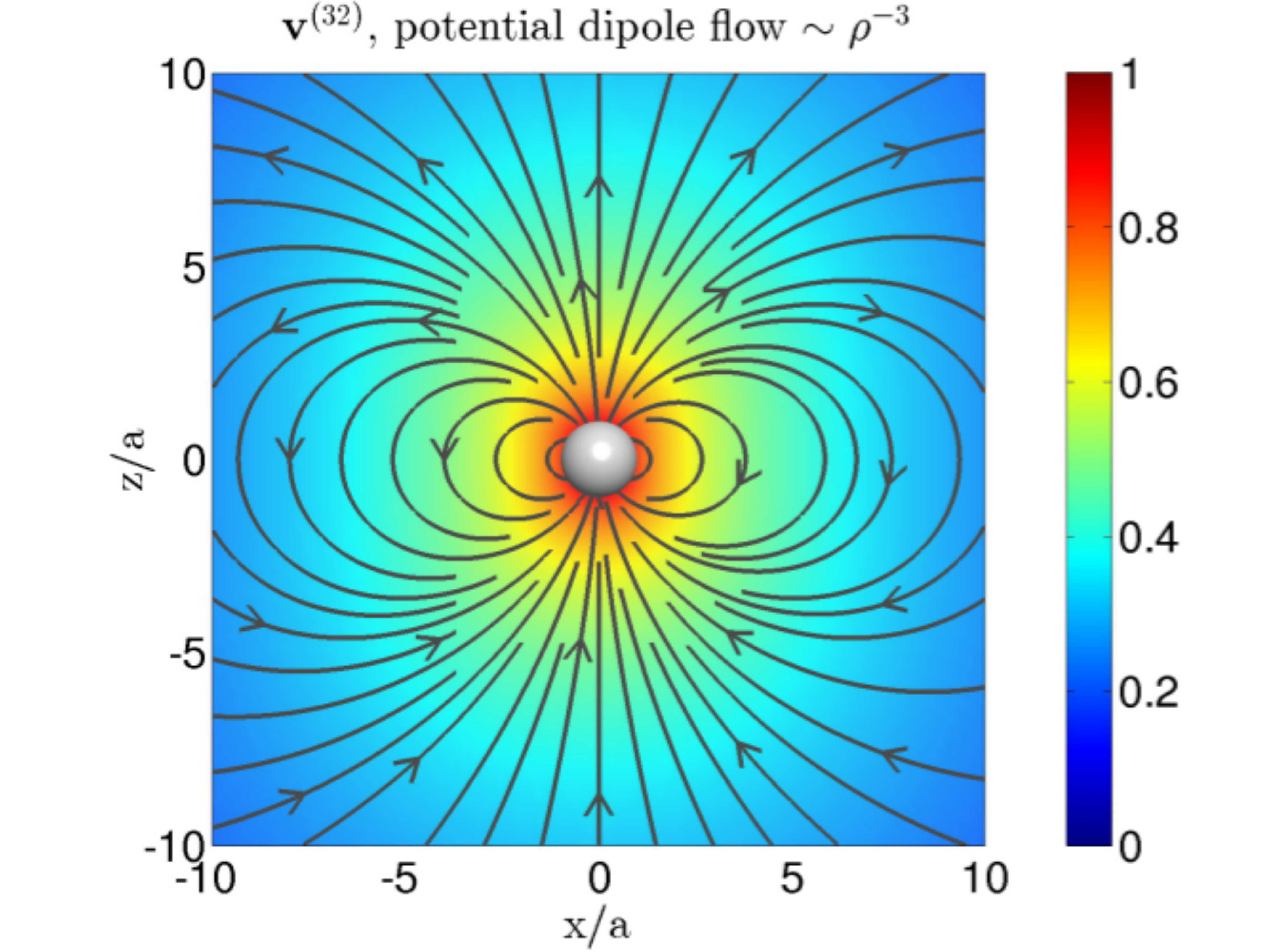}\\
\includegraphics[width=0.496\textwidth]{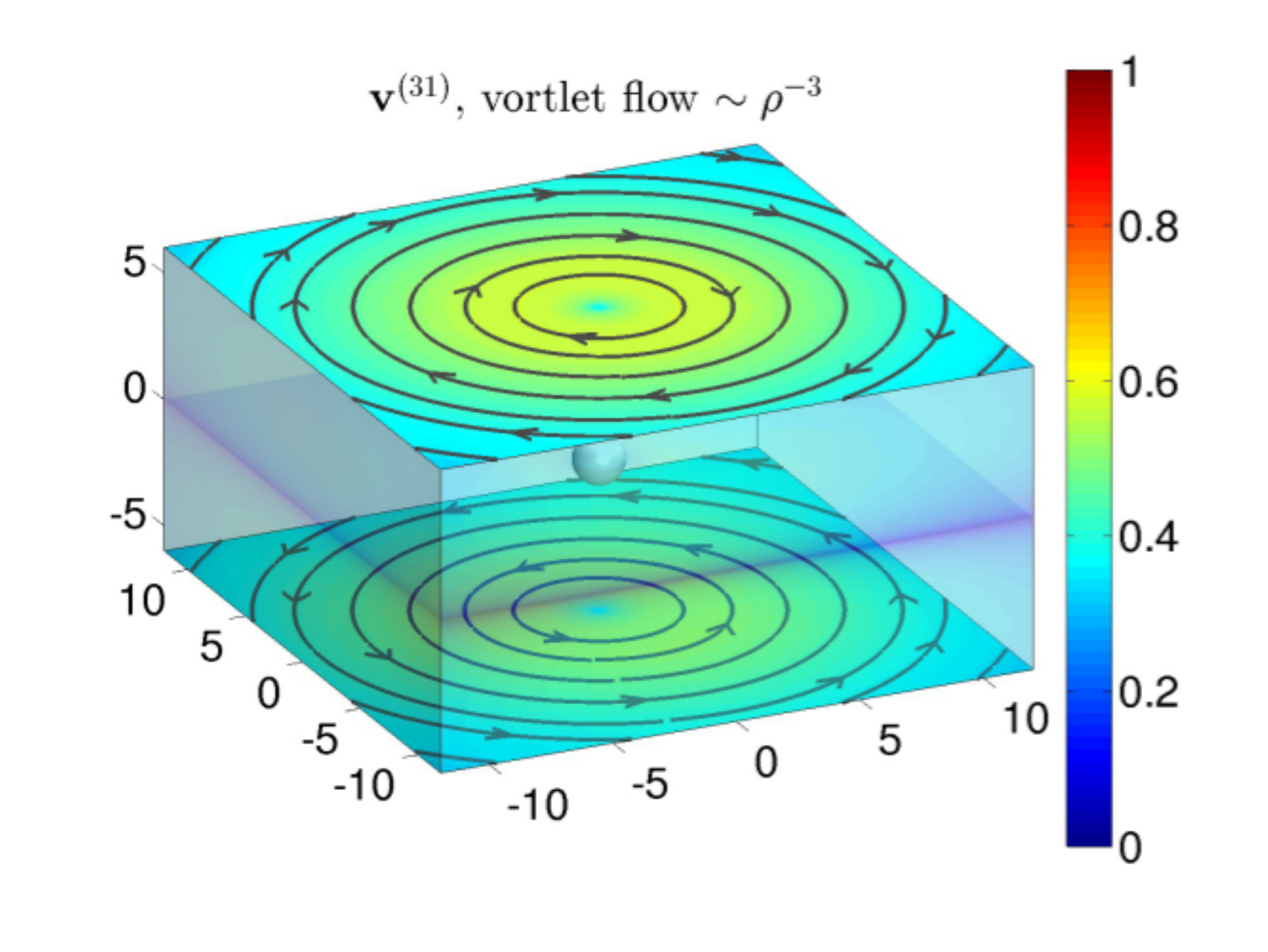}
\includegraphics[width=0.496\textwidth]{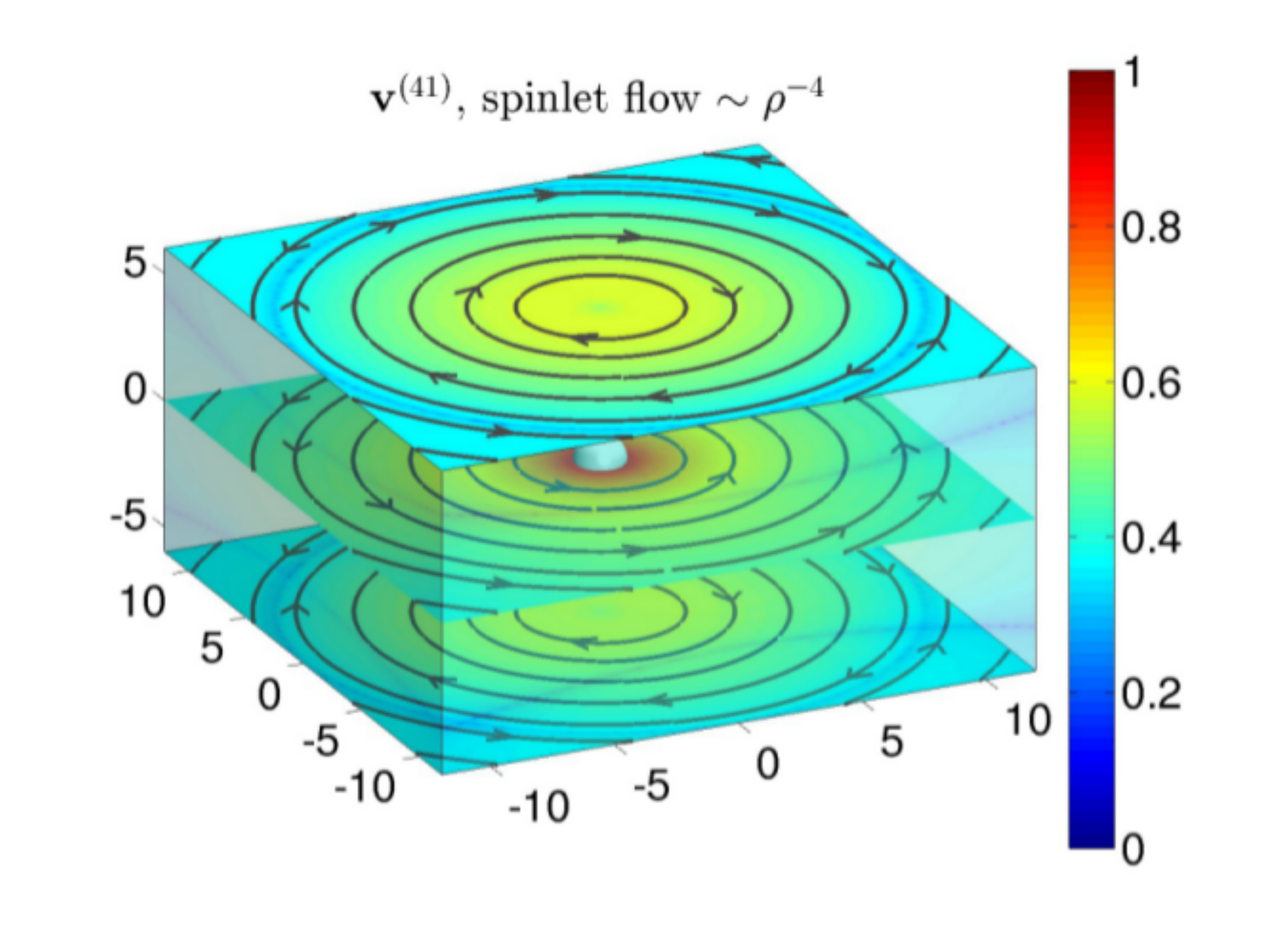}
\caption{
    Axisymmetric and swirling flows due to traction and velocity modes on the
    surface of a sphere.  Streamlines are overlaid on a pseudocolor plot of the
    logarithm of the magnitude of the fluid velocity in a planes containing the
    axis of the symmetry in the first four figures and in planes normal to it in
    the last two figures. The vectorial and septorial quadratic modes produce
    force-free, torque-free translation.  The cubic spinlet mode produces
    force-free, torque-free rotation.
    }
\label{fig:single-particle-flows}
\end{center}
\end{figure}

The structure of the irreducible components of the flow are shown explicitly for
$l=0, 1, 2, 3$ in table \ref{table:irred_fluid_flow}, together with the
contribution from a general $l$. From this table, we see that only $\sigma = 0$
terms have finite-sized corrections. Further, the finite-size correction for any
$l$ has the same form as the $\sigma = 2$ contribution from $l+2$. This pattern
is clearly seen at each order in the table. The flows corresponding to $l=0,1,
2, 3$ are plotted in figure \ref{fig:single-particle-flows} and discussed
further in the next section.

\begin{table}
  \begin{center}
  \begin{tabular}{ | c | c | c | c | }
  \hline
    $8\pi\eta\,\mf v^{(l\sigma)}$       & $\sigma=0$            & $\sigma=1$           & $\sigma=2$ \\ \hline
% ============================================================================
    &  & & 
    \\
    $l=0$      
    & 
    $%\left(1+\frac{a^{2}}{6}\nabla^{2}\right)
    \mathcal F_m^0\mf G \cdot\mf{Q}^{(10)}$
    &---  
    & --- 
    \\
    &  & & 
    \\
    \hline
% ============================================================================
% ============================================================================
    &  & & 
    \\
    $l=1$          & 
    $
    %\left(1+\frac{a^{2}}{10}    \nabla^{2}\right)
    \mathcal F_m^1\bm\nabla \mf G
    \odot\mf{Q}^{(20)}$
    &  $  -\frac{1}{2}  \big(\bm\nabla \times\mf G \big)  \cdot   \mf{Q}^{(21)} $    &---
    \\
    &  & & 
\\\hline
% ============================================================================
    &  & & 
    \\
    $l=2$      
    & 
    $%\left(1+\frac{a^{2}}{14}\nabla^{2}\right)
    \mathcal F_m^2\bm\nabla \bm\nabla  \mf G \odot\mf{Q}^{(30)}$
    &  $  -\frac{2}{3} \bm\nabla\big(\bm\nabla \times\mf G \big)\odot\mf{Q}^{(31)}$
    &  $\frac{2}{5}  \nabla^2 \mf G   \odot \mf{Q}^{(32)}   $
    \\
    &  & & 
    \\
    \hline
% ============================================================================
    &  & & 
    \\
    $l=3$      
    & 
    $%\left(1+\frac{a^{2}}{18}\nabla^{2}\right)
    \mathcal F_m^3\bm\nabla \bm\nabla \bm\nabla  \mf G 
    \odot\mf{Q}^{(40)}$
    &  $  -\frac{3}{4} \bm\nabla\bm\nabla\cdot\big(\bm\nabla \times\mf G
    \big)\odot\mf{Q}^{(41)}$
    &  $\frac{36}{35}  \bm\nabla\nabla^2 \mf G   \odot \mf{Q}^{(42)}   $
    \\
    &  & & 
    \\\hline
% ============================================================================
    &  & & 
    \\
    $\vdots$ 
    & $\vdots$       
    & $\vdots$       
    & $\vdots$      
    \\
    &  & & 
    \\ \hline
% ============================================================================
    &  & & 
    \\
    $l$          & 
    $% \left(1+\frac{a^{2}}{4l+6}\nabla^{2}\right) 
    \mathcal F_m^l\bm\Delta^{(l+1)}\bm\nabla^{(l)} \mf G   \odot   \mf{Q}^{(l0)}$
    &  $  -\frac{l}{l+1}\bm\Delta^{(l)}  \left(\bm\nabla^{(l)} \times\mf G \right)
    \odot\mf{Q}^{(l1)}$
    &  $\frac{l(l+1)}{2(2l+1)} \bm\Delta^{(l)} \bm\nabla^{(l-2)}\nabla^2 \mf G 
    \odot\mf{Q}^{(l2)}$
    \\
    &  & & 
    \\ \hline
% ============================================================================
  \end{tabular}
  \caption{
      Fluid flow due to a sphere in a Stokes flow consists of three terms at any
      order $l$, given by $\sigma=0,~1, \text{and } 2$. We use the fact the $\mf
      Q^{(l)}$ can be broken into three parts, $\mf Q^{(l\sigma)}$ which are
      individually symmetric and traceless in their last $l-\sigma$ indices.
      The fluid flow can then be written in terms of the Green's function of the
      Stokes flow which manifestly ensures incompressibility and biharmonicity
      (see Appendix~\ref{appendix:solution_green_function}). 
      }
  \label{table:irred_fluid_flow}
\end{center}
\end{table}
% ============================================================================
%
%
% ============================================================================
% ============================================================================
%
%
%

The system of linear equations that relate the velocity and traction coefficients are
\begin{align} \label{eq:linear_VFV_relation_yl}
   4 \pi \eta \,
  \mf{V}_n^{(l+1)}&
  =   -
  \sum_{m=1}^N
  \sum_{l' = 0}^{\infty}\left[
  \bm{G}_{nm}^{(l+1, \, l'+1)}(\mf{R}_n ,\, \mf{R}_m)
  \odot\mf{F}_m^{(l'+1)} 
  - \eta  \bm{K}_{nm}^{(l+1, \, l'+1)}(\mf{R}_n,\, \mf{R}_m)
  \odot\mf{V}_m^{(l'+1)}\right],
\end{align}
where the matrix elements are given by double integrals of the Greens' function
and the stress tensor,
\begin{subequations}
\label{eq:matrix_elemets_yl}
\begin{align}\label{eq:appendix_matrix_element_integral_single_layer}
\bm{G}_{nm}^{(l+1,\,l'+1)}
%(\mf{R}_n ,\, \mf{R}_m) 
&= 
    \frac{(2l+1)(2l'+1)}{(4\pi a^2)^2}\,  
    \int {\mf Y}^{(l)} (\bm{\rho_n}) \, \mf{G}(\mf R_n + \bm\rho_n, \mf{R}_m + \bm{\rho_m})  \,
    {\mf Y}^{(l')} (\bm{\rho_m}) \, \mrm{d} S_m \mrm{d} S,
\end{align}
\begin{align}\label{eq:flow_elements_appendix_double}
\bm{K}_{nm}^{(l+1,\, l'+1)}
%(\mf{R}_n ,\, \mf{R}_m) 
&= 
    \frac{2l+1}{4\pi a^2(l'-1)!(2l'+1)!!}\,  
    \int \mf Y^{(l)}(\bm{\rho_n}) \mf{K}(\mf R_n+\bm\rho_n, \mf R_m +
    \bm \rho_m)\cdot\mf n  \,
    \mf Y^{(l')}(\bm{\rho_m}) \, \mrm{d} S_m \mrm{d} S_n. 
\end{align}
\end{subequations}
As we show in Appendix \ref{appendix:matrix_elements}, these matrix elements can
be expressed in terms of \emph{derivatives} of the Greens' functions and the
stress tensor when the particle indices are not equal. For equal particle
indices, the Green's function matrix element is reduced to an angular
integration and the stress tensor matrix element is the identity tensor. These
results are collected below, 
\begin{subequations}
\begin{align}\label{eq:single_layer_yl_matrix_element}
\bm{G}_{nm}^{(l+1,\,l'+1)}&=   
    \left\{ 
  \begin{array}{l l l }
        \delta_{ll'}
        \,\frac{(2l+1)}{(2\pi a)}\,
        \int d\Omega\,
        \mf Y^{(l)}(\unv{\bm\rho})
        \left(\mathbb I - \unv{\bm\rho} \unv{\bm\rho}\right)
        \mf Y^{(l)}(\unv{\bm\rho}),
    \quad & m=n
   \\\\
        a^{l+l'} 
        \mathcal{F}_m^l
        \mathcal{F}_m^{l'}
        \,
        \bm\nabla_n^{(l )}\,
        \bm\nabla_m^{(l')}
        \,\mf G(\mf R_n,\, \mf R_m )
    ,
    \quad & m\neq n,
    \end{array} \right.
\end{align}
\begin{align}\label{eq:double_layer_yl_matrix_element}
\bm{K}_{nm}^{(l+1,\,l'+1)}&=   
    \left\{ 
  \begin{array}{l l l }
- \delta_{ll'}\,4\pi\mathbb I\, \bm\Delta^{(l)},
    \quad & m=n
   \\\\
     \frac{4 \pi a^{(l+l'+1)}}{(l'-1)!(2l' +1)!!}
        \mathcal{F}_m^l
        \mathcal{F}_m^{l'}
        \,
    \bm \nabla_n^{(l)}
    \bm \nabla_m^{(l'-1)}
        \,\mf K(\mf R_n,\, \mf R_m )
    ,
    \quad & m\neq n.
    \end{array} \right.
\end{align}
\end{subequations}
We note that the diagonal ($m=n$) expression of the $\bm K^{(l+1,\, l'+1)}_{nm}$
is defined for $l'\geq 1$ and $\bm K^{(l+1,\, l'+1)}_{nm} = 0$ for $l'=0$. We
emphasize that the diagonal matrix elements are evaluated using the
translational invariance of the Green's function and the stress tensor. For the
off-diagonal matrix elements, no such property is used, and the results,
therefore, are valid for any bounding geometry. 

The matrix elements having been determined, the rigid body motion can now be
obtained from the formal solution of the previous section, in terms of the
mobility and propulsion matrices, 
(\ref{eq:galerkin_representation_of_linear_VF_relation_yl}),

\begin{subequations}\label{eq:genralized_active_stokes_law}
\begin{align}
    \mf{V}_n &= \mf{V}_n^{a} +
  \sum_{m}^N
    \Big[
    \bm\mu^{\mrm{TT}}_{nm}\cdot\mf F^e_m +
    \bm\mu^{\mrm{TR}}_{nm}\cdot\mf T^e_m  
    \Big]+ 
  \sum_{m\neq n}^N
  \sum_{l = 1}^{\infty}
  \left[
  \bm\pi_{nm}^{(\mrm T,\, l+1)}\odot \mf V_m^{(l+1)} 
  \right],
    \\
  \bm{\Omega}_n &= \bm{\Omega}_n^{a} +
  \sum_{m}^N
    \Big[
\bm\mu^{\mrm{RT}}_{nm}\cdot\mf F^e_m +
\bm\mu^{\mrm{RR}}_{nm}\cdot\mf T^e_m 
    \Big]+ 
\sum_{m\neq n}^N   \sum_{l = 1}^{\infty}
  \left[
  \bm\pi_{nm}^{(\mrm R,\, l+1)}\odot \mf V_m^{(l+1)} 
  \right].
\end{align}
\end{subequations}
The active velocity $\mf V^a_n$ and active angular velocity $\bm\Omega_n^a$ are 
\begin{subequations}
\begin{align}
  4\pi a^2 \, \mf{V}_n^{a}     &=- \int\mf{v}^a(\mf{r}_n) \,\mrm dS_n, \\
  4\pi a^2 \,\bm{\Omega}^{a}_n &=-
  \frac{3}{2a}\int\unv{\bm{\rho}}_n\times\mf{v}^a(\mf{r}_n)\,\mrm dS_n,
\end{align}
\end{subequations}
the mobility matrices are
\begin{subequations}
\label{eq:mobility_matrices}
\begin{align}
8\pi\eta\,\bm\mu_{nm}^{\mrm{TT}} &=
\delta_{nm} \bm G^{(1,\, 1)}_{nn} +
        \left( 1-\delta_{nm} \right)
        \Big(
         \bm G^{(1,\, 1)}_{nm}
        -\big[\mathsf G^{\mrm{LH}}
        (\mathsf G^{\mrm{HH}})^{-1}
         \mathsf G^{\mrm{HL}}\big]_{nm}^{(1,\,1)}
        \Big),
\\
8\pi\eta\,\bm\mu_{nm}^{\mrm{TR}} &=\frac12\bm\epsilon:
        \left( 1-\delta_{nm} \right)
        \Big(
         \bm G^{(1,\, 2)}_{nm}
        -\big[\mathsf G^{\mrm{LH}}
        (\mathsf G^{\mrm{HH}})^{-1}
         \mathsf G^{\mrm{HL}}\big]_{nm}^{(1,\,2)}
        \Big),
\\
8\pi\eta\,\bm\mu_{nm}^{\mrm{RT}} &=\frac12\bm\epsilon:
        \left( 1-\delta_{nm} \right)
        \Big(
         \bm G^{(2,\, 1)}_{nm}
        -\big[\mathsf G^{\mrm{LH}}
        (\mathsf G^{\mrm{HH}})^{-1}
         \mathsf G^{\mrm{HL}}\big]_{nm}^{(2,\,1)}
        \Big),
\\
8\pi\eta\,\bm\mu_{nm}^{\mrm{RR}} &=
\frac14\delta_{nm} \bm G^{(2,\, 2)}_{nn} +\frac14
        \left( 1-\delta_{nm} \right)
        \Big(
         \bm G^{(2,\, 2)}_{nm}
        -\big[\mathsf G^{\mrm{LH}}
        (\mathsf G^{\mrm{HH}})^{-1}
         \mathsf G^{\mrm{HL}}\big]_{nm}^{(2,\,2)}
        \Big),
\label{}
\end{align}
\end{subequations}
and the propulsion matrices are
\begin{subequations}
\label{eq:propulsion_matrices}
\begin{align}
8\pi\,\bm\pi_{nm}^{(\mrm T,\, l+1)} &=
        \bm K^{(1,\, l+1)}_{nm} + 
    \big[
         \mathsf G^{\mrm{LH}}
        (\mathsf G^{\mrm{HH}})^{-1}
        (4\pi\,\mathbb{I} -\mathsf K^{\mrm{HH}})
        \big]^{(1,\,l+1)}_{nm},
\\
8\pi\,\bm\pi_{nm}^{(\mrm R,\, l+1)} &=
%\frac{1}{2}\bm\epsilon:
         \bm K^{(2,\, l+1)}_{nm}+
    \big[
         \mathsf G^{\mrm{LH}}
        (\mathsf G^{\mrm{HH}})^{-1}
        (4\pi\,\mathbb{I} -\mathsf K^{\mrm{HH}})
        \big]^{(2,\,l+1)}_{nm}.
\label{}
\end{align}
\end{subequations}
The above expressions are generalizations of Stokes law, which expresses the linearity
of tractions and velocities, to the motion of $N$ active particles and are central results
of this section. 

It is useful to compare these results with existing results for computing
many-body hydrodynamic interactions of spheres. In the absence of active
velocities the boundary condition reduces to the usual no-slip boundary
condition on the surface of each sphere. Then, there is no contribution from the
second integral of the boundary integral representation and we obtain a
single-layer formulation, first used by Zick and Homsy \cite{zick1982stokes},
for computing the mobility of a periodic suspension. We should emphasise that
Zick and Homsy used a reducible polynomial basis, in which redundant
polynomials have to be manually removed from the sum. In contrast, the
tensorial spherical harmonics are irreducible and, therefore, do not contain
any redundant terms. In this limit of vanishing active velocity, truncating
the Galerkin expansion to linear terms yields the method of computing
far-field hydrodynamic interactions in the so-called ``FTS'' Stokesian
dynamics method of Brady and colleagues \cite{durlofsky1987dynamic,
brady1988dynamic}. This low-order truncation has been subsequently extended
upto to $7$-th order in polynomials by Ichiki \cite{ichiki2002improvement}.
However, Ichiki's extension requires six separate steps to relate the
velocity coefficients to the traction coefficients. In contrast, the method
provided here, since it uses identical basis functions for both velocity and
traction, directly provides expressions for the coefficient matrices and the
problem is reduced, directly, to solving the linear system. The extension of 
Stokesian dynamics method to active particles and its comparison to the method 
presented here, has been been done in section \ref{section:discussions}.

Returning to the case of active velocities, several authors have used the method of
reflections for computing hydrodynamic interactions between pairs of
spheres.  Keh and Chen \cite{chen1988electrophoresis, keh1995particle}
consider both the electrophoretic and thermophoretic interactions between
particle pairs, computed to $\mathcal{O}(\rho^{-7} )$ in particle separation $\rho$. 
Anderson \cite{anderson1981concentration} has computed the change in
the electrophoretic mobility in an infinite suspension as a function of
volume fraction using the superposition approximation. Rider and O'Brien
\cite{rider1993dynamic} computed the AC electrophoretic mobility as a
function of volume fraction and frequency of applied field, again in the
superposition approximation. 
%Ishikawa et al. \cite{ishikawa2006,ishikawa2008coherent, koch2011} have
%considered spheres with axisymmetric slip velocities, truncated to the first
%two non-trivial contributions, and computed the far-field contribution to
%the rigid body motion in the superposition approximation, while using a
%lubrication approximation to compute the near-field contribution. In
%contrast, our work includes both axisymmetric and swirling components of the
%active velocity and does not make any separation of far-field and near-field
%hydrodynamic interactions. Ishikawa et al. also use boundary element method,
%that is, a collocation discretization of the single layer boundary integral
%equation, to compute near-field hydrodynamic interactions.
None of these
papers have recognised, though, that hydrodynamic interactions between
active particles can be completely described by propulsion matrices, nor
have any of them provided a recipe for their evaluation, both of which have
been accomplished in this paper. 
%%
%%
%===========================================================================
\section{Minimal truncation and superposition approximation} \label{section:minimal-active-spheres}
% ============================================================================
%%
In this section, we consider a truncated version of the theory developed in the
previous section, that retains the essential aspects of the hydrodynamic
interactions between active spheres. Our truncation retains the least number of
terms that are necessary to produce force-free translations and torque-free
rotations of a single active spheres and to include all long-ranged
contributions to the hydrodynamic interaction between many active spheres. It
is, in this sense, a minimally truncated version of the general theory of the
previous section. As we now show, expanding the boundary fields to cubic order
in surface polynomials, that is retaining terms up to $\mf Y^{(3)}(\unv{\bm
\rho})$, provides the desired minimal truncation. 

In this minimal truncation, only terms corresponding to $l = 0, 1, 2, 3$ in
(\ref{eq:VF_Yl_expansion}) are retained
\begin{subequations}
\begin{align}
4\pi a^2\, \mf f(\mf R + \bm \rho) = 
                          \Big[\mf F^{(1)} 
                            + 3\,\mf F^{(2)} \odot \mf Y^{(1)}(\unv{\bm\rho}) 
                            + 5\,\mf F^{(3)} \odot\mf Y^{(2)}(\unv{\bm\rho}) 
                            + 7\,\mf F^{(4)}\odot \mf
                            Y^{(3)}(\unv{\bm\rho})
                          \Big],
\end{align}
\begin{align}
\mf v(\mf R + \bm \rho) = \Big[\mf V^{(1)} 
                            + \mf V^{(2)} \odot \mf Y^{(1)}(\unv{\bm\rho}) 
                            + \frac16\,\mf V^{(3)} \odot\mf Y^{(2)}(\unv{\bm\rho}) 
                            + \frac{1}{90}\,\mf V^{(4)}\odot \mf
                            Y^{(3)}(\unv{\bm\rho})
                          \Big].
\end{align}
\end{subequations}
The reducible traction coefficients are expressed in terms of their irreducible parts as 
(\ref{eq:irred_traction_coefficients}) 
\begin{subequations} \label{appeq:stress_multipole_embeddings}
\begin{align}
\mf F^{(1)}     &=  - \mf F^e, 
\\
a\, \mf F^{(2)} &= \mf S + \frac12 \bm\epsilon\cdot\mf T^e,
\\
a^2\, \mf F^{(3)} &= \mf \Gamma -\frac23
\bm\Delta^{(2)}\odot\left(\bm\epsilon\cdot\mf \Psi\right) + \frac35
\bm\Delta^{(2)} \odot\bm\delta\mf D,
\\
a^3\, \mf F^{(4)} &= \mf \Theta 
-\frac34\bm\Delta^{(3)}\odot\left(\bm\epsilon\cdot\mf \Lambda\right)
+ \frac67\bm\Delta^{(3)} \odot\bm\delta\mf \Xi,
\end{align}
\end{subequations}
while the reducible velocity coefficients are, similarly, expressed in terms of
their irreducible parts as 
(\ref{eq:irred_velocity_coefficients})
\begin{subequations} \label{appeq:velocity_multipole_embeddings}
\begin{align}
\mf V^{(1)}     &= \mf V - \mf V^a, 
\\
\frac1a\, \mf V^{(2)} &= \bm s - 
\bm\epsilon\cdot(\mf\Omega-\mf\Omega^a),
\\
 \frac{1}{a^2}\, \mf V^{(3)} &= \bm \gamma -\frac23
\bm\Delta^{(2)}\odot\left(\bm\epsilon\cdot\bm \psi\right) + \frac35
\bm\Delta^{(2)} \odot\bm\delta\bm d,
\\
\frac{1}{a^3}\, \mf V^{(4)} &= \bm \theta 
-\frac34\bm\Delta^{(3)}\odot\left(\bm\epsilon\cdot\bm \lambda\right)
+ \frac67\bm\Delta^{(3)} \odot\bm\delta\bm \xi.
\end{align}
\end{subequations}
Here, $\mf F^e$, $\mf T^e$ and $\mf S$ are the familiar force, torque and
stresslet strengths. $\mf D$, $\mf \Psi$ and $\mf\Gamma$ are the irreducible
parts of the quadratic coefficients $\mf F^{(3)}$. They are irreducible tensors
of rank 1, 2, and 3 as expected from angular momentum algebra. Similarly, $\mf
\Theta$, $\mf\Lambda$ and $\mf\Xi$ are the irreducible parts of the cubic
coefficients $\mf F^{(4)}$, and again, are tensors of rank 2, 3 and 4
\cite{coope1965,jerphagnon1970, jerphagnon1978, andrews1982}. The respective
velocity coefficients are written in lower case. The flow expression
(\ref{eq:flow_expression}) for a single particle for each $l,\,\sigma$ is
given in table~(\ref{table:irred_fluid_flow}). The coefficients are linear 
combinations of the traction and velocity coefficients, to four of which we give
explicit names 
\begin{subequations}
\begin{align}
\mf Q^{(20)} &=  \mf F^{(20)} - \frac{8\pi\eta a}{3}\, \mf V^{(20)}  &&\text{(Stresslet)},\\
\mf Q^{(30)} &=  \mf F^{(30)} - \frac{8\pi\eta a}{15}\, \mf V^{(30)} &&\text{(Septlet)},\\
\mf Q^{(31)} &=  \mf F^{(31)} - \frac{16\pi\eta a}{45}\, \mf V^{(31)} &&\text{(Vortlet)},\\
%\mf Q^{(32)} =  \mf F^{(32)} - 12\pi \eta a\, \mf V^{(32)} &&&\text{(potential
%dipole)}.\\
\mf Q^{(41)} &=  \mf F^{(41)} -\frac{\pi\eta a}{35} \mf V^{(41)} &&\text{(Spinlet)}.
\end{align}
\end{subequations}
Recalling the definition of the expansion coefficients, (\ref{eq:expansion_coefficients}), we find that 
\begin{align}  \label{eq:single_sphere_Laddyzhenskaya_Stresslet}
    a\mf{Q}^{(20)} 
    =&
    \int \left[
    \frac12\left(\mf{f}^{} \, {\bm{\rho}} 
    + \left( \mf{f}^{} \, {\bm{\rho}} \right)^{{T}}\right)
    -{\eta}\,\left(
    \mf{v}^{} \, \widehat{\bm{\rho}} 
    + \left( \mf{v}^{} \, \widehat{\bm{\rho}} \right)^{{T}}\right)
    \right]\mrm dS,
\end{align}
which, as we mentioned earlier, is the stresslet \cite{batchelor1970} introduced by
Landau and Lifshitz \cite{landau1987fluid}. The direct formulation of the boundary integral
equation used here yields this combination in a natural manner and clearly shows
the origin of each of the contributions from the Green's function and the stress
tensor integrals. It also provides a natural extension of this combination for
expansion coefficients for any value of $l$ and $\sigma$. The three
combinations explicitly named in the above, are particularly useful for the
minimal description of spheres with active boundary layers. Their significance
is as follows. The septlet is a third rank irreducible tensor and its flow
decays as $1/\rho^3$. The vortlet and spinlet are, respectively, second rank and
third rank irreducible tensors. They are, respectively, the leading and next to
leading order terms that produce torque-free swirling flows. All these flows are
plotted in figure \ref{fig:single-particle-flows} when the expansion tensors are uniaxial. 

The irreducible traction and velocity coefficients are related by the solution
of the diagonal part of (\ref{eq:linear_VFV_relation_yl}), and on using
(\ref{eq:single_layer_yl_matrix_element}) and
(\ref{eq:double_layer_yl_matrix_element}), we find
that the relation is both diagonal and
scalar, 
\begin{align}
8\pi \eta\, \mf V_n^{(l\sigma)} = \delta_{ll'}\delta_{\sigma\sigma'}\,
G_{nn}^{(l\sigma,\, l\sigma)}
\mf F_n^{(l'\sigma')}.
\end{align}
Explicitly, for each irreducible part we have
\begin{subequations} \label{eq:single_sphere_traction_velocity_relations}
  \begin{align}
  \label{eq:single_sphere_Force_torque}
\mf{V} &=\mf V^a + \frac{\mf F^e} { 6\pi\eta\,a}
     ,\qquad
~\qquad
\mf{\Omega} =\mf \Omega^a +  \frac{\mf T^e} { 8\pi\eta\,a^3}
\\  \label{eq:single_sphere_stress_septlet}
\mf S
    &=- \frac{20\pi\eta\,a^3}{3}\mf s, 
\qquad
\, ~ ~ \qquad\bm \Gamma
    =- \frac{7\pi\eta\,a^5}{6}\mf \gamma ,
\\  \label{eq:single_sphere_vortlet}
\bm \Psi
    &= \frac{4\pi\eta\, a^5}{3}\bm \psi ,\qquad
~ ~ ~ ~ ~ \qquad\mf D
    =- 2\pi\eta\,a^5\mf d ,
\\  \label{eq:single_sphere_spinlet}
\bm \Lambda
    &=- \frac{2\pi\eta\,a^7}{15}\bm \lambda .\qquad
%
%
%8\pi\eta a^3\, \bm \xi
%    =- \frac{4}{a^4}\mf \Xi ,
%%
 \end{align}
\end{subequations}
The first of these equations shows that translation and rotation are possible in the
absence of external forces and torques. In their absence, the strengths of the potential
dipole and the spinlet are related to the active translation and rotational velocities. This
is evident from the expressions of these coefficients given in Appendix (\ref{appendix:solution_green_function}). 
%%
%% 
% ============================================================================
\begin{figure}
\begin{center}
\includegraphics[width=\textwidth]{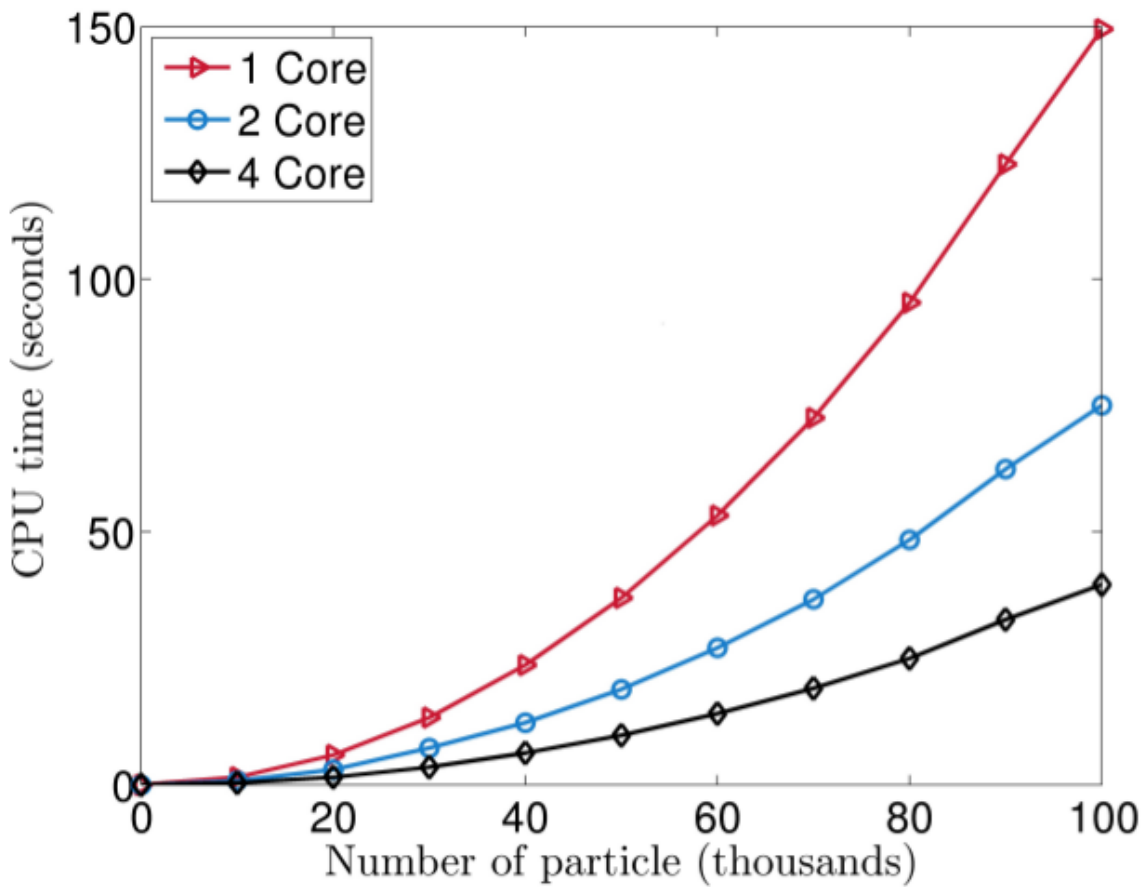} 
\caption{
    Benchmarks for calculation of stresslet propulsion matrix $\pi^{\mrm{Ts}}$
    in the superposition approximation on a 4-core machine using the PyStokes
    library \cite{pystokes}. The present implementation of the library shows a
    linear scaling with the number of cores and quadratic scaling with the
    number of particles.
    }
\label{fig:benchmarks}
\end{center}
\end{figure}
% ============================================================================
%
%
At each order $l$, there are $(2l+3) + (2l+1) + (2l - 1)$ coefficients, from
each of the three irreducible parts at that order. This gives $3$ coefficients
at $l=0$, which are the three components of the force, $8$ coefficients at
$l=1$, which are the $5$ components of the stresslet and $3$ components of the
torque, $15$ coefficients at $l=2$, which are the $7$ components of the septet,
$5$ components of the vortlet, and $3$ components of the potential dipole, and
$21$ components at $l=3$. The linear system at this order of truncation, then,
is of size $47N\times47N$. For particles that number between tens to hundreds of
thousands, this leads to a large numerical system that must be solved
efficiently. The full details of the numerical implementation, using matrix-free
iterative solvers and fast methods for matrix vector products, will be presented
in a future contribution. 

A significant simplification occurs at small concentrations, when the distance
between particles is large compared to their radius. At small volume fractions,
the mean distance between active particles is large compared to their size. The
off-diagonal terms in the linear system, which depend on the distance between
particles, are therefore small. Consequently, all off-diagonal terms in 
(\ref{eq:single_layer_yl_matrix_element}, \ref{eq:single_layer_yl_matrix_element})
can be neglected, and the boundary integral equation becomes diagonal in the
particle indices. The values of the tractions obtained from the one-body
solution are, then, a good approximation for the many-body situation. At this
order of approximation, the off-diagonal matrix elements in the mobility and
propulsion matrices are negligible, and only the direct superposition
contribution remains. This superposition approximation to the hydrodynamic
interaction was first utilised by Kirkwood and Riseman \cite{kirkwood1948intrinsic} in their study of polymer
dynamics. With this approximation, the mobility matrices are
\begin{subequations}
\label{eq:mobility_matrices}
\begin{align}
8\pi\eta\,\bm\mu_{nm}^{\mrm{TT}} &=
        \mathcal{F}_n^0
        \mathcal{F}_m^0 \mf G,~ ~ ~
\qquad
8\pi\eta\,\bm\mu_{nm}^{\mrm{TR}} =\frac12
        \bm\nabla_m\times\mf G,
\\
8\pi\eta\,\bm\mu_{nm}^{\mrm{RT}} &=\frac12
        \bm\nabla_n\times\mf G,
\qquad
8\pi\eta\,\bm\mu_{nm}^{\mrm{RR}} =
    \frac14\bm\nabla_n\times\left(\bm\nabla_m\times\mf G\right),
\label{}
\end{align}
\end{subequations}
and the propulsion matrices are
\begin{subequations}
\label{eq:propulsion_matrices}
\begin{align}
8\pi\eta\,\bm\pi_{nm}^{\mrm {Ts}} &=
        C_{20}\,\mathcal{F}_n^0
        \mathcal{F}_m^1\bm\nabla_m \mf G, \qquad~ ~~ ~ ~
8\pi\eta\,\bm\pi_{nm}^{\mrm {Rs}} =
       \frac{C_{20}}{2} \bm\nabla_n\times(\bm\nabla_m \mf G),
\\
8\pi\eta\,\bm\pi_{nm}^{\mrm {T\gamma}} &=
        C_{30}\,\mathcal{F}_n^0
        \mathcal{F}_m^2\bm\nabla_m\bm\nabla_m \mf G,\qquad
8\pi\eta\,\bm\pi_{nm}^{\mrm {R\gamma}} =
        \frac{C_{30}}{2} 
        \bm\nabla_n\times
        (\bm\nabla_m\bm\nabla_m \mf G),\qquad
\\
8\pi\eta\,\bm\pi_{nm}^{\mrm{T\psi}} &=
        C_{31}\,
        \bm\nabla_m(\bm\nabla_m \times\mf G),\qquad ~ ~
8\pi\eta\,\bm\pi_{nm}^{\mrm {R\psi}} =\frac{C_{31}}{2}
        \bm\nabla_n\times(
        \bm\nabla_m(\bm\nabla_m \times\mf G)),
\\
8\pi\eta\,\bm\pi_{nm}^{\mrm{ Td}} &=
        C_{32}\,
        \nabla_m^2 \mf G,\qquad\qquad\qquad ~ 
8\pi\eta\,\bm\pi_{nm}^{\mrm{Rd}} =0
\\
8\pi\eta\,\bm\pi_{nm}^{\mrm{ T\lambda}} &=
        C_{41}\,
        \bm\nabla_m \bm\nabla_m(\bm\nabla_m \times\mf G), ~ ~ ~
8\pi\eta\,\bm\pi_{nm}^{\mrm{R\lambda}} =\frac{C_{41}}{2}
        \bm\nabla_n\times(\bm\nabla_m
        \bm\nabla_m(\bm\nabla_m \times\mf G)).
\label{}
\end{align}
\end{subequations}
The coefficients $C_{l\sigma}$ are derived using
(\ref{eq:single_sphere_traction_velocity_relations}). The specific 
values can be found in table \ref{tab:c-l-sigma}.
This level of approximation yields a very simple and direct method for 
studying active hydrodynamic interactions in dilute systems, through
a pairwise summation of the mobility and propulsion matrices. We have
implemented this superposition method, using a direct summation, in the
PyStokes library \cite{pystokes}.  In figure~\ref{fig:benchmarks}, 
we have plotted the time for computing the contribution from a 
 typical propulsion matrix, in this case the $l=1$ contribution.  
The library takes less than a minute for evaluating the velocity of
$\mathcal{O}(10^5)$ particles with prescribed stresslets. The method
scales quadratically in the number of particles and linearly in the number
of cores. Therefore, this provides method to compute the collective 
dynamics of a large number of hydrodynamically interacting active particles
on multi-core computational architectures. 
%%

% ============================================================================================
\section{Squirmers in a harmonic potential}\label{section:squirmers_trap}
% ============================================================================
Lighthill \cite{lighthill1952} and Blake \cite{blake1971a} considered a sphere
with a general axisymmetric slip velocity on the surface as a model for
ciliated organisms. Retaining only tangential surface flows and the first two
terms of their velocity expansion, the active velocity is 
\begin{subequations}
\begin{align}
v^a_{\rho} = \unv{\bm\rho}\cdot \mf v^a = 0, \qquad
v^a_{\phi} = \unv{\bm\phi}\cdot \mf v^a = 0, \qquad
v^a_{\theta} =\unv{\bm\theta}\cdot \mf v^a = B_1 \sin\theta + B_2 \sin\theta \cos\theta.
\label{}
\end{align}
\end{subequations}
which can be written more compactly as, 
\begin{align}
\mf v^a =  \left(\unv{\bm \rho} \unv{\bm\rho} - \mathbb I\right)\cdot 
\left[
B_1\mf p + B_2 \left(\mf p\mf p -\frac{\mathbb I}{3}\right)\cdot\unv{\bm\rho}
\right],
\end{align}
where $\mf p$ is the orientation vector and $B_1$ and $B_2$ are constants.
In terms of tensorial spherical harmonics, this is 
\begin{align}
\mf v^a = 
-\frac{2B_1}{3} \mf p
- \frac{3B_2}{5} \left(\mf p\mf p -\frac{\mathbb I}{3}\right)
\cdot \mf Y^{(1)}(\unv{\bm\rho})
+ \frac{B_1}{3}\mf p\mathbb I \odot\mf Y^{(2)}(\unv{\bm\rho})
+\frac{B_2}{15} \left(\mf p\mf p -\frac{\mathbb I}{3}\right)\mathbb I 
\odot\mf Y^{(3)}(\unv{\bm\rho}).
\label{eq:squirmer-active-velocity-expansion}
\end{align}
Comparing (\ref{eq:squirmer-active-velocity-expansion}) with the velocity
expansion in tensorial spherical harmonics
(\ref{eq:velocity-expansion-spherical-harmonics}), we note that $l=1,2,3$
contribute to the flow created by the squirmer. Also, the specific form of the
coefficients of $\mf Y^{(l)}$ ensures that only 
$\mf V^{(20)}$, $\mf V^{(32)}$, $\mf V^{(42)}$ have non-vanishing contributions. These are%
\begin{subequations}
\begin{align}
\mf V^{(10)} &= \mf V -\mf V_n^a,\qquad 
\mf V^{(20)}  = -\frac35 B_2\left(\mf p\mf p -\frac{\mathbb I}{3}\right), \\
\mf V^{(32)} &= \frac{15}{2}\mf V^a_n,\qquad \quad
\mf V^{(42)}  = \frac{140}{3}B_2\left(\mf p\mf p -\frac{\mathbb I}{3}\right) ,
\end{align}
\end{subequations}
where the active translational velocity $\mf V^a_n$ of the squirmer is 
\begin{align}
\mf V^a_n = -\langle \mf v^a \rangle = \frac{2 B_1}{3} \mf p .
\label{}
\end{align}
This is an example of a non-trivial solution of the traction, which satisfies
the zero-force, zero-torque constraint of
(\ref{eq:newtons_equations_force_torque_free_constraints}), and yet, produces particle
translation. The exterior flow due this boundary velocity consists of a
stresslet, a potential dipole, and a degenerate octupole. 
Also, once the velocity coefficients are known then the tractions coefficients
can be calculated using (\ref{eq:single_layer_yl_matrix_element}) and
(\ref{eq:single_sphere_traction_velocity_relations}).
\begin{align}
4\pi a^2\,\mf f = 
4\pi\eta a B_2\left(\mf p\mf p -\frac{\mathbb I}{3}\right) 
\cdot \mf Y^{(1)}(\unv{\bm\rho})
-12\pi\eta a   B_1\mf p\cdot \mf Y^{(2)}(\unv{\bm\rho})
-\frac{560}{3}\pi\eta a B_2 \left(\mf p\mf p -\frac{\mathbb I}{3}\right)   
\odot\mf Y^{(3)}(\unv{\bm\rho}).
\label{eq:squirmer-active-velocity-expansion}
\end{align}
So the flow expression due to the squirmer is a linear combination of the flow
due to these three terms as per table~(\ref{table:irred_fluid_flow})
\begin{align}
8\pi\eta\,\mf v  =
    C_{20}\mathcal F_m^1\bm\nabla\mf G\odot\mf Q^{(20)}
    +C_{32}\nabla_n^2\mf G\odot\mf Q^{(32)}
    +C_{42}\bm\nabla\nabla^2\mf
    G\odot\mf Q^{(42)}.
\label{}
\end{align}
%
%%
% ============================================================================
\begin{figure}
\begin{center}
\includegraphics[width=\textwidth]{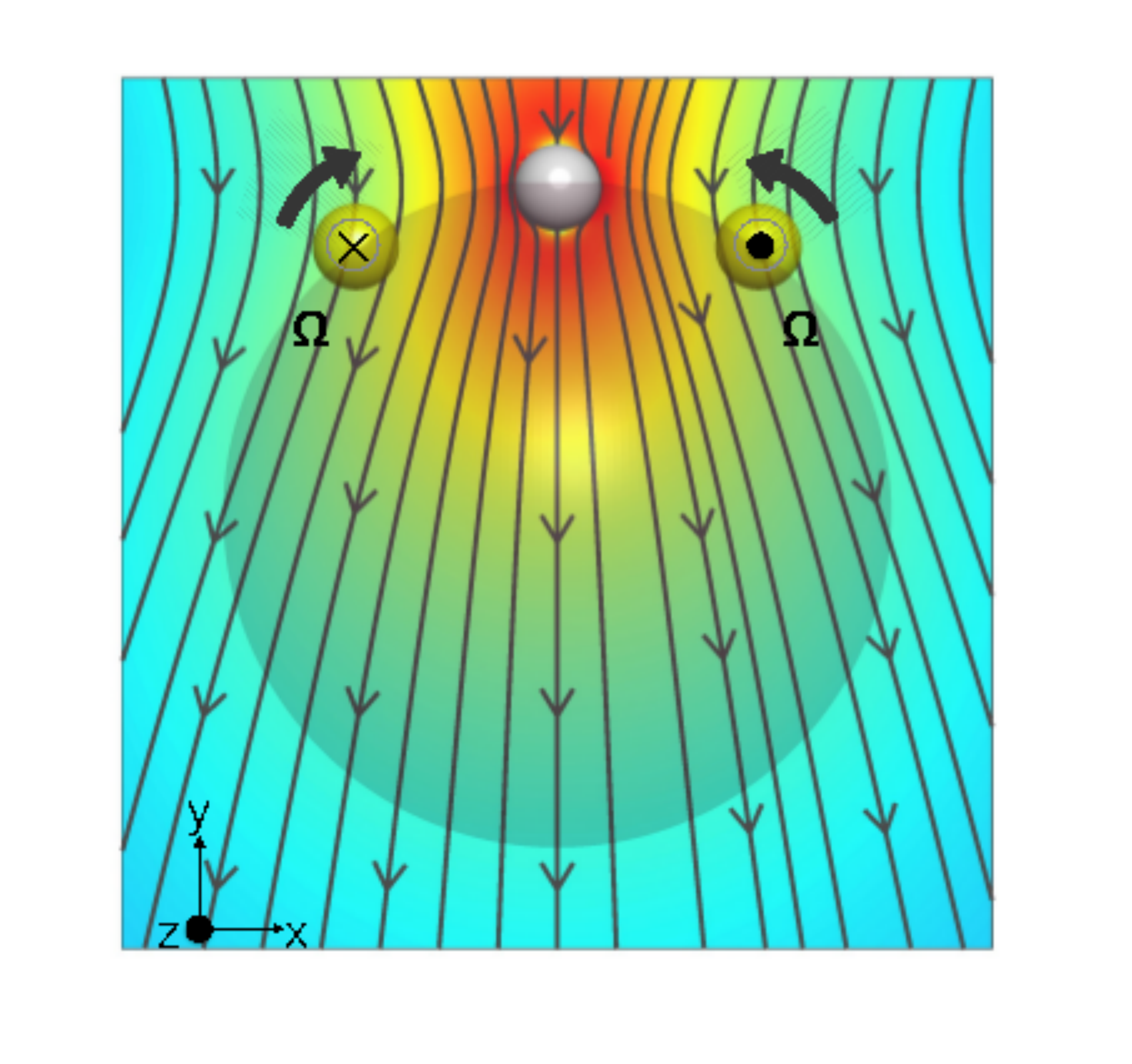} 
\caption{Flow generated by a test sphere (white) with a body force acting on it in the downward direction. 
This is the Stokeslet flow due to the particle at the north pole.
The particles are situated on the surface of the schematic sphere. 
The two tracer particles (yellow ones) on the surface of the sphere 
then swim towards the test particle by orientating along it.} 
\label{fig:stokeslet_flow}
\end{center}
\end{figure}
%% ============================================================================
%
%%============================================================================
Active suspensions under confinement produces interesting dynamical phases 
\cite{nash2010, woodhouse2012, hennes2014self, ezhilan2015transport}. 
Here we study effects of confinement on hydrodynamically interacting finite sized squirmers of radius $a$. 
To do a minimal modeling of the squirmers, we truncate the velocity and angular velocity 
update equation (\ref{eq:genralized_active_stokes_law}) at $l=2$. The equations of the motion
of the squirmers in the external harmonic potential is then
\begin{subequations}\label{eq:trap_eom}
\begin{align} \label{eq:trap_position}
  \dot{\mf{R}}_n
  =& 
    \underbrace{\mu^{\mrm{TT}}_{nn}}_{\mathcal O(1)} (-\bm\nabla_n U)
   +\sum_{m\neq n}^N\bigg[
    \underbrace{\bm\mu^{\mrm{TT}}_{nm}}_{\mathcal O(\rho^{-1})} \cdot (-\bm\nabla_m U) 
    +\underbrace{\bm\pi_{nm}^\mrm{{Ts}}}_{\mathcal O(\rho^{-2})} \odot\bm s_m
    +\underbrace{\bm\pi_{nm}^\mrm{{Td}}}_{\mathcal O(\rho^{-3})} \odot\bm d_m
    \bigg]
    +\underbrace{\mf V^a_n}_{\mathcal O(1)},
\end{align}
\begin{align} \label{eq:trap_position}
  {\mf{\Omega}}_n
  =& 
   \sum_{m\neq n}^N\bigg[
    \underbrace{\bm\mu^{\mrm{RT}}_{nm}}_{\mathcal O(\rho^{-2})} \cdot (-\bm\nabla_m U) 
    +\underbrace{\bm\pi_{nm}^{\mrm{Rs}}}_{\mathcal O(\rho^{-3})} \odot\bm s_m
    \bigg],
\end{align}
\end{subequations}
where $U$ is the potential due to the external force on each particle and the active velocity 
$\mf V^a_n = v_s \mf p_n$, where $v_s$ is the squirmer speed. 
We use the following uniaxial parametrization for the stresslet $\bm s$ and potential dipole $\bm d$,
\begin{align}
\bm s = s_0 \left(\mf p \mf p -\frac13 \mathbb I\right),\qquad
\bm d = d_0 \mf p,
\end{align}
where $ s_0$ and $d_0$ are stresslet and potential dipole strengths respectively. 
Since the particles are
harmonically confined, the force due to the potential is the usual spring force 
$\mf F_n = -\bm\nabla_n U = -k\mf R_n$, where $k$ is the stiffness of the trap. 

The equation describing the motion of a single passive particle in harmonic trap 
of spring constant $k$ can be written as,
\begin{align}
\mf{\dot{R}} = \mu_{nn}^{\mrm{TT}} \mf F^e = \frac{\mf{F^e}}{6\pi\eta a} = -\frac{k\mf R}{6\pi\eta a}.
\end{align}
%
%%% ============================================================================
\begin{figure}
\begin{center}
\includegraphics[width=\textwidth]{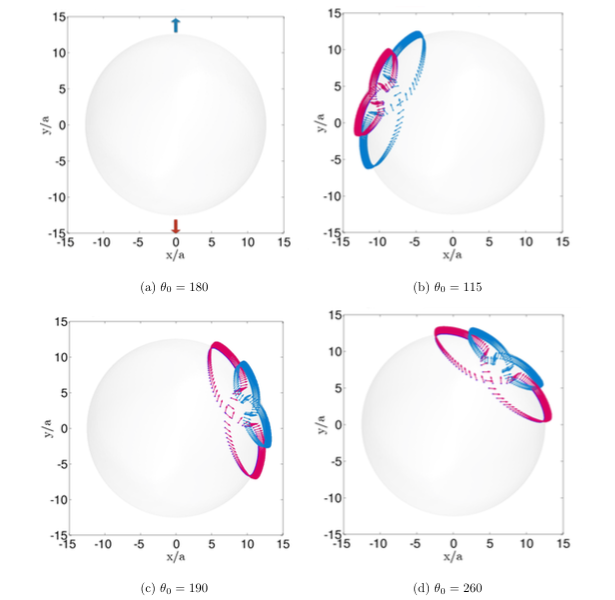}
%\subfigure[~$\theta_0 = 180$]{\includegraphics[width=0.45\textwidth]{Figure5a}} 
%\subfigure[~$\theta_0 = 115$]{\includegraphics[width=0.45\textwidth]{Figure5b}} 
%\subfigure[~$\theta_0 = 190$]{\includegraphics[width=0.45\textwidth]{Figure5c}} 
%\subfigure[~$\theta_0 = 260$]{\includegraphics[width=0.45\textwidth]{Figure5d}} 
%%%
\caption{Dynamics of two particles in a harmonic trap. The particles are initialised on the 
surface of the sphere, with radius $R^*=\mathcal A a = 12.56 a$ (indicated by a schematic sphere), 
which is their stable state 
without HI. The first particle (blue) is initialised at the north pole and the second particle is kept at 
an angle $\theta_0$ measured in anticlockwise direction from the position of the first particle. 
In panel (a) $\theta_0=180$ and hence
the angular velocity (\ref{eq:harmonicTrap_Omega}) of each particle is identically zero and hence there is 
no dynamics.
In panel (b) $\theta_0 =115$ and particles form a closed 8-like orbit.
In panel (c) and (d), the final orbit is plotted for $\theta_0$ taking the values 190 and 260. It can
be thus seen that two particle dynamics depends critically on the initial condition. 
The axis of the orbit, thus formed, is also dependent on the initial conditions.
}
\label{fig:two-particles}
\end{center}
\end{figure}
%% ============================================================================
The given dynamical equation for the particle is solved exactly by, 
$\mf{R}(t) = \mf{R}_0 e^{-t/\tau}$.
So, there is a time scale, $\tau_1$ in the system, given by
${\tau}_1 = \frac{6\pi\eta a}{k}$.
This is the time scale due to the overdamped motion of the particle in a harmonic potential.
The final stable state is where all the particles are near the origin of the harmonic
potential. If the particle is active, there is an additional term 
due to the active velocity of the particle
$\mf{\dot{R}} = -\frac{k\mf R}{6\pi\eta a} + v_s \mf{p}$.
The stable state of this system is found by setting the right hand side to zero, which defines a 
radius of confinement given by,
$R^* = \frac{ 6\pi\eta v_s}{k} a = \mathcal Aa$,
such that at the confinement radius, the net radial velocity of the particle is zero. 
The stable configuration is now inverted and hence in the long time limit all the particle 
are found on the surface of the sphere of radius $R^*$ 
\cite{tailleur2008statistical, tailleur2009sedimentation, nash2010}.
So a system of the non-interacting squirmer will have a stable state on the surface of the
sphere of the radius $R^*$. We use this as the initial condition for our simulations 
and study the evolution of this stable initial state once the hydrodynamic interactions 
are turned on.

The lowest order terms in the velocity update equation (\ref{eq:trap_eom}) are the self terms 
$\mu^{\mrm{TT}}_{nn}(-\bm\nabla_n U) $ and $\mf V^a_n$. The lowest 
order term from the HI (hydrodynamic interactions) is the 
Stokeslet $\bm\mu^{\mrm{TT}}_{nm}$ and decays inversely $\rho^{-1}$, as 
the distance between the particles increases. In figure \ref{fig:stokeslet_flow}, the flow due to a 
Stokeslet on the north pole of the schematic sphere, is plotted. 
It is shown that the two other particles, which are 
moving in the field of the Stokeslet on north pole, move towards the source sphere by
orienting towards it. This observation is crucial for the motion of particles in the
harmonic potential, as Stokeslet is the most dominant term in the velocity and angular 
velocity equations once a large number of particles are considered.

Here we discuss simplest possible treatment of the HI in the trap. Since we are working in the 
superposition approximation, the HI is pairwise additive, hence the two particles dynamics should 
give us insight for problem of many squirmers in the harmonic potential. Consider the $\mathcal O (1)$
terms in the position update equation, we see that the stable position is the surface of the sphere, which
we take as the initial condition. So unless we rotate the particle towards the trap center, nothing 
interesting will happen. The angular equations then determine the stability of this state. 
The leading terms in the angular velocity equation are 
$\bm\mu_{nm}^{\mrm{RT}}$ and
$\bm\mu_{nm}^{\mrm{Rs}}$
as the potential dipole does not contribute to the angular velocity.
%
%% ============================================================================
\begin{figure}[!t]
\begin{center}
\subfigure[\,t=$0$       ]{\includegraphics[width=0.496\textwidth]{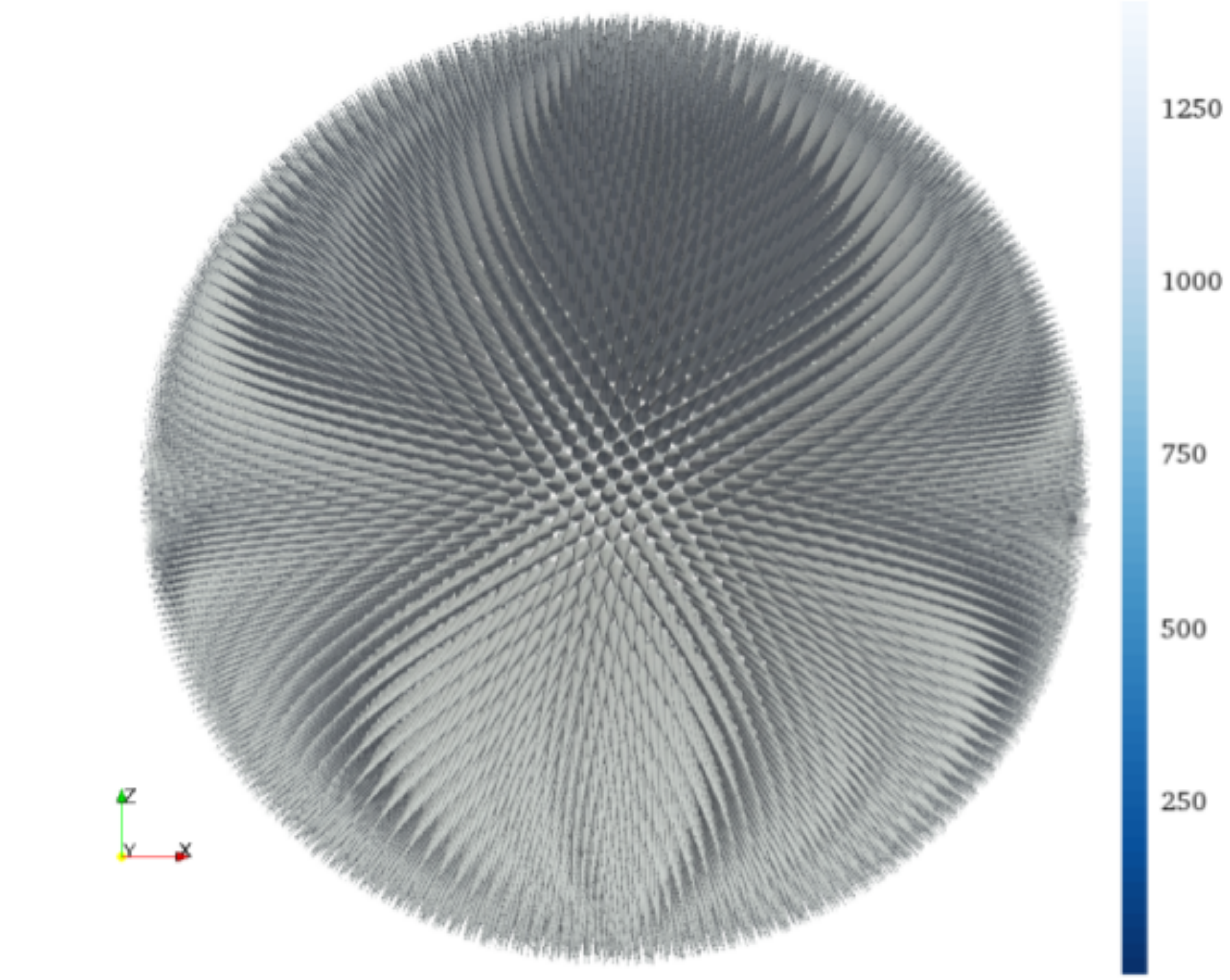}} 
\subfigure[\,t=$\tau_r $ ]{\includegraphics[width=0.496\textwidth]{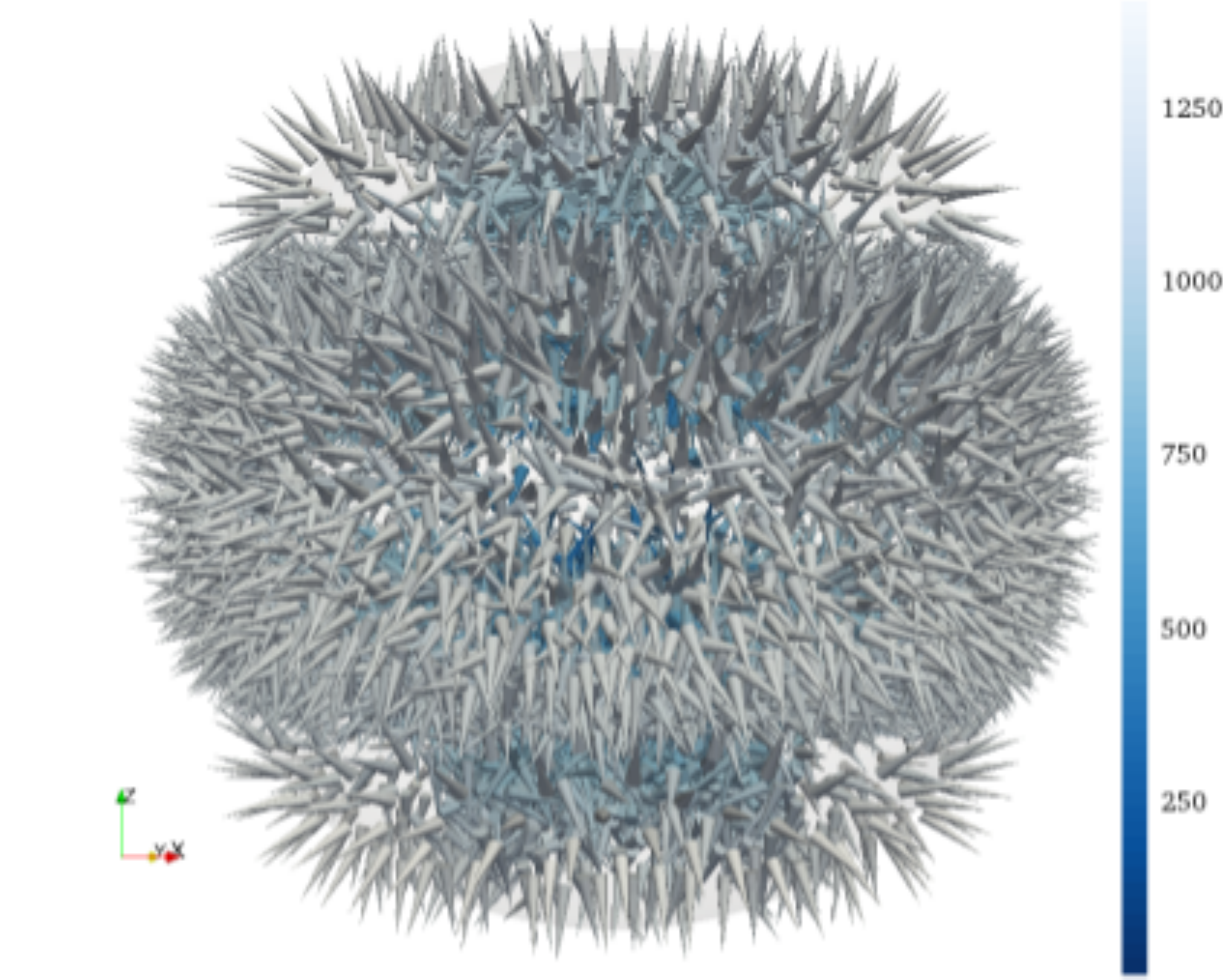}} 
\subfigure[\,t=$2\tau_r $]{\includegraphics[width=0.496\textwidth]{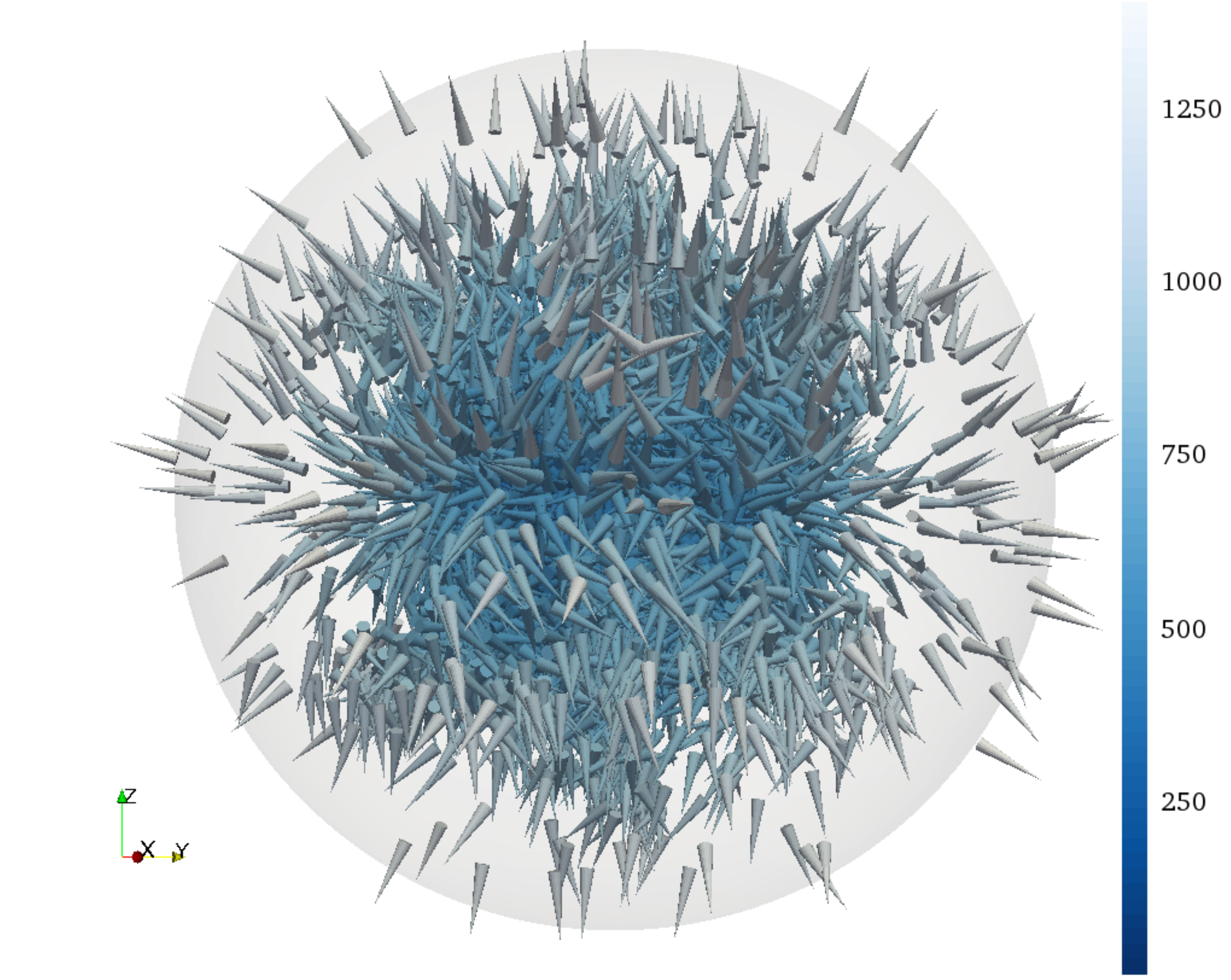}} 
\subfigure[\,t=$8\tau_r $]{\includegraphics[width=0.496\textwidth]{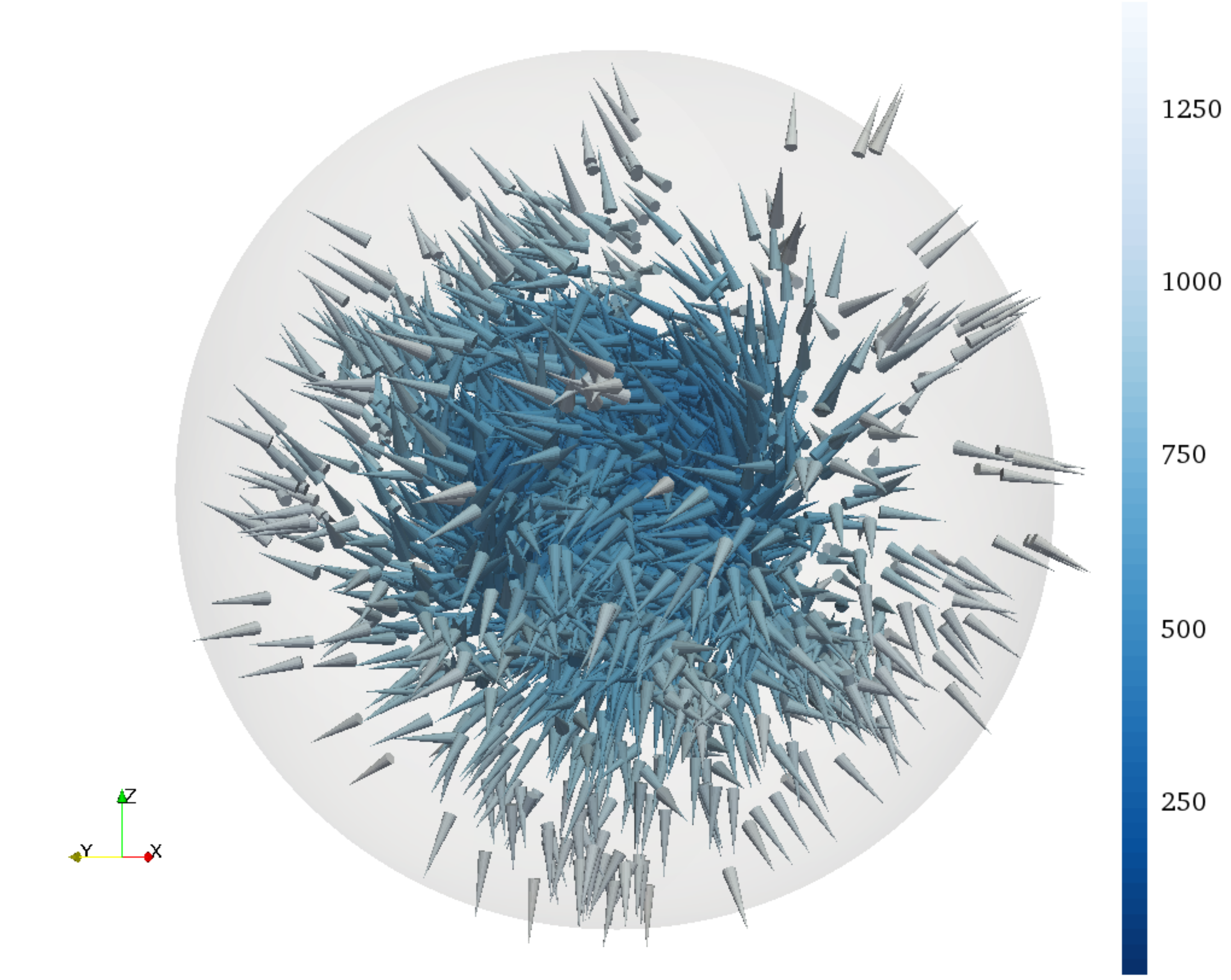}} 
\caption{
Snapshots from the simulation of $10^4$ active particles, confined in a harmonic
potential, at different times in terms of the rotational time scale $\tau_r =
\frac{8\pi\eta a}{k}$.  Squirmers are spherical particles with a radius $a$ and
orientation vector $\mf p$. In the figures, the squirmers are denoted by cones
located at the particle position and pointing in the direction of their
orientation vector.  The colormap shows the distance of the particles
from the origin.
The initial condition at time $t=0$ corresponding to a uniform distribution of the squirmers 
on the surface of trap sphere of radius $R^*=\mathcal A a=1256.6 a $ (indicated
by a schematic sphere). With time, steady-state convective currents are
established and, the so-called self-assembled pump is formed in the system.
See the supplementary material for the Movie. 
The simulations were performed using the PyStokes library \cite{pystokes}.}
\label{fig:N-particles}
\end{center}
\end{figure}
%% ============================================================================
%
For two particles the orientational dynamics is then given by
\begin{subequations}\label{eq:harmonicTrap_Omega}
\begin{align}
\mf{\dot{p_{1}}} &= \left(\frac{-k\,(\mf{R_{2}}\times \mf{R_1})}{8\pi\eta \rho^3} + 
		 \frac{3C_{20}s_{0}\,(\mf p_2 \cdot \bm\rho)\, (\mf{p_{2}}\times \bm{\rho}) }{\rho^5}  \right)\times\mf p_1,
\end{align}
\begin{align}
\dot{\mf{p_{2}}} &=\left( \frac{-k\,(\mf{R_{1}}\times \mf{R_2})}{8\pi\eta \rho^3} 
                + \frac{3C_{20}s_{0} \,(\mf p_1 \cdot \bm\rho)\,(\mf{p_{1}}\times \bm{\rho})}{\rho^5}  
                \right)\times\mf p_2.
\end{align}
\end{subequations}
Thus the angular velocity depends on the coordinates of other particles 
and not their orientations, at lowest order. This dominant term sets the rotational 
time scale in the system $\tau_r = \frac{8\pi\eta a}{k}$. 

The dynamics of the two particles depends critically on their initial condition. If two particles are 
initialised on two diametrically opposite points then the there is no contribution to the angular
velocity as $\bm\mu_{nm}^{\mrm{RT}}$ and $\bm\mu_{nm}^{\mrm{Rs}}$ are identically zero and hence the particle do not rotate. Therefore the dynamics is frozen. If the angle between particles 
is different from 180 degrees, then they form an 8-like structure which is a closed orbit. 
In figure \ref{fig:two-particles} we have plotted the closed orbits formed by a system of two particles. 
The two-particle dynamics is then strongly determined by the initial condition. As the 
angular velocity (\ref{eq:harmonicTrap_Omega}) of each particle is identically zero for 
$\theta_0 =180$, there is no rotation for this initial condition. For any other initial angle 
between the particles, a closed orbit is formed. The axis of this orbit is dependent on the initial 
condition of the particles as can be seen in figure \ref{fig:two-particles}.

In figure \ref{fig:N-particles}, snapshots from the simulation of $10^4$ finite-sized active 
particles confined in 
a harmonic potential have been given at different times in terms of the rotational time 
scale $\tau_r = \frac{8\pi\eta a}{k}$, determined by the Stokeslet contribution to the 
angular velocity in (\ref{eq:trap_eom}). The initial condition at $t=0$ corresponding to a 
uniform distribution of the squirmers on the surface of the sphere of radius
$R^*=\mathcal A a=1256.6 a $. 
This corresponds to the stable distribution of active squirmers without HI. 
We see the destabilization of this structure through the reorientation induced by the HI. 
The system eventually obtains a steady-state of convective currents, the so-called 
self-assembled pump \cite{nash2010}. 
Eq. (\ref{eq:trap_eom}) show that the key ingredients necessary for the pumping state are 
(a) one-body self-propulsion and 
(b) two-body rotation from the vorticity of the Stokeslet induced in the trap.
The interesting things happen near the surface of the particles when the particle can not 
go any further, as the radial velocity is zero and need a angular velocity to rotate them 
back. On account of the hydrodynamic interactions from other particles, 
the particles on the surface are rotated 
back to the center of the trap. The dynamics is strongly determined by the 
Stokeslet flow as given in figure \ref{fig:stokeslet_flow}. The particles 
tends to come together and move towards the center of the trap but as they 
approach the center their orientations are rotated and then the one-body 
active velocity terms start dominating in making them move back to 
the surface of the confining sphere. Again, as they move towards
the confining sphere, the Stokeslet strength picks up and pulls the particle back. This 
results in a pump-like motion of a macroscopic number of particles inside the trap. 
We also see that the two particles form a closed orbit about the surface of the trap 
and hence in the $N$ particle limit the behavior should be qualitatively similar, 
which is indeed the case. Now since the number of particles is large, they tend 
to bring the particles more closer to the origin till the Stokeslet contribution are 
weak and the particles again start moving back to the surface of the confining sphere. 
We note that only the leading order terms in (\ref{eq:trap_eom}) are important and 
account for the self-assembled pump. Thus the necessary and sufficient condition to 
obtain the pumping state are one-body self-propulsion and two-body rotation 
from the vorticity of the Stokeslet induced in the trap. 
%%
%%%
%%%%

\section{Discussion and summary}\label{section:discussions}
In the preceding sections, we have developed a systematic theory for studying
hydrodynamic interactions in active colloidal particles. It is instructive to
compare our approach with existing descriptions of active matter. These can be
broadly classified into kinematic theories that prescribe active motion, without
considering the balance of mass, momentum, angular momentum and energy, and
dynamic theories which derive the active motion from the balance of these
conserved quantities. The models of Toner and Tu \cite{toner1995long}, Vicsek
\cite{vicsek1995novel}, and Chate \cite{gregoire2004onset} belong the former
category, while those of Finlayson and Scriven \cite{finlayson1969}, Simha and
Ramaswamy \cite{simha2002a}, and Saintillan and Shelley \cite{saintillan2008a}
belong to the latter category \cite{ramaswamy2010, marchetti2013,
cates2012diffusive}. 
The models,
both kinematic and dynamic, can be also be classified by the length and time
scales at which they resolve matter. Hydrodynamic theories operate at the
coarsest length and time scales, and retain only variables that relax slowly.
Kinetic theories operate at smaller length and time scales and contain in them
the hydrodynamic description. Finally, particulate models offer a scale of
resolution higher than both of hydrodynamic and kinetic theories and offer a
complementary description free of the requirement of a continuum limit. 

In the context of the above classification, our approach is a momentum-conserving
particulate model for active matter. Active motion in our approach is not prescribed
by fiat, but appears as a consequence of the balance of forces and torques, both at
the particle boundaries and in the bulk ambient fluid. The contact forces and torques at
the boundary are supplied by non-equilibrium processes that occur in the boundary
layer which appear, in our approach, as an active velocity. We parametrize this velocity
in full generality, and thus, all possible forms of active motion that arise from boundary
layer phenomena are, in principle, included in our description. The specificity is contained
in the velocity expansion coefficients, which depend on additional fields like electrical or
chemical potentials in physico-chemical contexts, or on the motility of organelles, like cilia, 
in biological contexts. Thus, our approach provides a generic framework for active matter 
without sacrificing specific detail. 

The Stokesian dynamics method has been extended to active particles, as reviewed in the 
work of Koch and Subramanian \cite{koch2011}. In particular, Ishikawa et al. 
\cite{ishikawa2006, ishikawa2008development} have 
considered spheres with axisymmetric slip velocities, truncated to the first
two non-trivial contributions, and computed the far-field contribution to
the rigid body motion in the superposition approximation, while using a
lubrication approximation to compute the near-field contribution. In
contrast, our work includes both axisymmetric and swirling components of the
active velocity and does not make any separation of far-field and near-field
hydrodynamic interactions. Ishikawa et al. also use boundary element method,
that is, a collocation discretization of the single layer boundary integral
equation, to compute near-field hydrodynamic interactions.

Momentum conservation in our approach can be enforced without the need for explicit
fluid degrees of freedom. This is possible at low Reynolds numbers, as the momentum
balance equation for the fluid reduces to an elliptic partial differential equation whose
solution can be represented as integrals over the domain boundaries. In contrast to models
that retain explicit fluid degrees of freedom to enforce momentum conservation
\cite{ramachandran2006lattice, nash2010, nash2008, jayaraman2012,
delmotte2015large, llopis2006dynamic}, our approach eliminates
explicit fluid degrees of freedom and yet retains momentum conservation.
Momentum conservation is enforced at the boundaries and is automatically ensured
in the bulk through the integral representation. This reduction of the
three-dimensional partial differential equation to a two-dimensional integral
equation leads to efficient numerical methods
for dynamic simulations of active particles. 

Squirmers have been studied previously in the work of Pagonabarraga and Llopis \cite{llopis2006dynamic}
using lattice Boltzmann (LB) methods. They subtract a constant amount of momentum with fixed
magnitude from the fluid, in a solid cone, which lie in predefined direction of motion of 
the particle. The subtracted momentum is then added to the active particle, such that momentum
conservation is ensured. In another work Baskaran and Marchetti \cite{baskaran2009statistical} 
model the active swimmer by a asymmetric rigid dumbbell composed of two Stokeslet differing in size,
and hence the leading order hydrodynamic interaction in their case is $\mathcal{O}(\rho^{-2})$. 
They have then used a multipole expansion and a continuum analysis of the model. In contrast, we 
assume a spherical particle with active velocity specified on its surface which is then expanded in 
a Galerkin basis which leads to a systematic series of term which has a nice group-theoretic classification.
And hence any generic mechanism generating the active velocity can be modeled in our scheme of 
expansion in a complete orthonormal Galerkin basis.

In this paper, we have provided explicit expression for the boundary integrals and matrix elements
for a spherical active particles in an unbounded fluid. The extension to other geometries is accomplished by using a
Green's function appropriate to that geometry. Thus, active colloidal
suspensions near plane walls and in periodic domains can be treated straightforwardly within this
method. The periodic domain needs special care as a naive flow summation
is conditionally convergent and a limiting procedure is needed to render the flow 
unconditionally convergent \cite{o1979method, brady1988dynamic}.

Hierarchical assemblies of active particles like single filaments with free
\cite{jayaraman2012} or clamped \cite{laskar2013} boundary condition and active
suspensions of short filaments \cite{pandey2014flow} can be studied under this
framework. Suspension rheology beyond the dilute can be studied systematically
within our approach. Extensions of the present study in the above
directions will be presented in future contributions. 

%

%% Acknowledgment
We thank S. Ambikasaran, M. E. Cates, P. Chaikin, G. Date, A. Donev, L.
Greengard, A. J. C. Ladd, I. Pagonabarraga, D. Pine, M. Shelley, H. A. Stone, and P. B. Sunil
Kumar for many useful discussions; M. Rao, H. Stark, S. Saha, and R. Winkler for
useful comments at the ICSM 2014 in Jaipur where this work was first presented;
M. E. Cates and A. Donev for constructive remarks on an earlier version of this
manuscript; and the Department of Atomic Energy, Government of India, for financial support. 

%%%%
%%%%
%%%%
%%%%
% ============================================================================================
% ============================================================================================
% ============================================================================================
% ============================================================================================
% ============================================================================================
\appendix
% ============================================================================================
\section{ Boundary integral equation for electrostatics }\label{appendix:electrostatic_analogy}
% ============================================================================================
The electrostatic problem of computing the potential due to $N$ spheres has a
similar form as (\ref{eq:many_body_boundary_integral_formulation}). 
Here, the potential is determined by Laplace's equation,
\begin{align} 
\nabla^2 \psi = 0 , 
\label{eq:laplace_equation}
\end{align}  
which has an integral representation \cite{jackson1962classical}
\begin{align} \label{eq:Laplace_equation}
  4 \pi\epsilon \, \psi(\mf{r})
  &=
   \sum_{m=1}^N \int \left[
  G(\mf{r},\mf{r_m} )\, \sigma(\mf r_m) - \epsilon \, n_i K_i(\mf r, \mf r_m)  \,\psi(\mf{r}_m)\, \right] \, \mrm{d} S_m,
  \quad 
  \substack{
  \displaystyle \mf{r} \in V 
  \\ \\ 
  \displaystyle \quad\,\mf{r} \in S_m
  }
%   }
\end{align}
\begin{align}
G(\mf{r},\mf{r}' ) = \frac{1}{\rho }, \qquad  
K_i(\mf{r},\mf{r}' ) =  \nabla_i G= -\frac{\rho_i}{\rho^3}.
\label{}
\end{align}
The potential in space, $\psi$, is
expressed as integrals over the particle surfaces of the potential and its normal derivative $\mf n
\cdot\bm\nabla \psi = \epsilon\sigma$, where $\epsilon$ is the permittivity and
$\sigma$ is the surface charge density. Thus, there is
a close analogy between the microhydrodynamic and electrostatic problems, with
the correspondence
\begin{align}
\mf v  \leftrightarrow \psi, \qquad   \mf f  \leftrightarrow \sigma, \qquad
\eta \leftrightarrow \epsilon.
\label{}
\end{align}
The analogy is not wholly complete since $n_i K_i$ is the normal derivative of
the Green's function of the Laplace equation, while $n_k K_{ijk}$ is the sum of
derivatives of the Green's function of the Stokes equation and the pressure
Green's function. However, it does provide a heuristic for understanding
microhydrodynamic phenomena guided by electrostatic analogies.
%%
%%
%%
%%
% ============================================================================================
\section{Evaluation of boundary integrals}\label{appendix:flow_elements}
% ============================================================================================
%
In this appendix, we outline the derivation of (\ref{eq:flow_expression}), from
(\ref{eq:flow-elements}), which expresses the boundary integral in terms of the
derivatives of the Green's function. The key idea is to Taylor expand the
Green's function about the center of the sphere, and express the $l$-th degree
polynomial of the radius vector in terms of tensorial spherical harmonics.
Orthogonality of the harmonics and biharmonicity of the Green's function reduces
this infinite number of terms in the Taylor series to exactly two, giving the
result in (\ref{eq:flow_expression}). To show these steps explicitly, the
Taylor expansion is 
\begin{align}\label{eq:G_expansion}
\mf G(\mf r,\, \mf R + \bm \rho)
               =& 
               \sum_{l=0}^{\infty}  \frac{1}{l!}(\bm\rho\cdot\bm\nabla)^{(l)}
               \,\mf G(\mf r, \, \mf R +\bm \rho) \Big|_{\bm \rho =0},
\end{align}
where the expansion of the symmetric $l$-th degree polynomial in tensorial spherical
harmonics is 
\begin{align}\label{}
(\bm\rho\cdot\bm\nabla)^{(l)}
=&
(\rho_{\alpha_1}\rho_{\alpha_2}\rho_{\alpha_3}\cdots\rho_{\alpha_l})
(
\nabla_{\alpha_1}
\nabla_{\alpha_2}
\nabla_{\alpha_3}\cdots
\nabla_{\alpha_l}).
\end{align}
It should be noted that 
$
(
\rho_{\alpha_1}
\rho_{\alpha_2}
\rho_{\alpha_3}\cdots
\rho_{\alpha_l})
$ is not in its irreducible form. Now to make use of the orthogonality relation
of (\ref{eq:tensorial_spherical_harmonics_orthogonality}) we convert the
$\bm \rho^{(l)}$ to tensorial spherical harmonics using the following trick \cite{mazur1982}
\begin{align}
\label{eq:mazur1982_trick}
\rho_{i_1}
\rho_{i_2}
\rho_{i_3}\cdots
\rho_{i_{l+2}}
=& a^l
\frac{Y^{(l)}_{i_1i_2\cdots i_{l+2}}}{(2l-1)!!}
+\frac{a^{l}}{2l-1}\sum_{\text{jk pairs}}\delta_{i_j i_k}
\frac{Y^{(l-2)}_{i_1i_2\cdots i_{j-1}i_{j+1} \cdots i_{k-1}i_{k+1}\cdots i_{l}}}
{(2l-5)!!}
+\mathcal{O}(l-4).
\end{align}
from which it follows that
\begin{align}
(\bm\rho\cdot\bm\nabla)^{(l)}
=&a^l
\left[
     \frac{\mf Y^{(l)}\bm\nabla^{(l)} }{(2l-1)!!}
    +\frac{1}{2l-1}\sum_{\text{jk pairs}}
    \frac{\mf Y^{(l-2)}\nabla^{2}\bm\nabla^{(l-2)}}{(2l-5)!!}
    +\mathcal{O}(l-4)
\right].
\end{align}
Thus, only two terms remain on integration, giving 
\begin{align}
\bm{G}^{(l+1)}(\mf{r} ,\, \mf{R}) &= 
    a^{l}\bm\Delta^{(l)}\left( 1 + \frac{a^2}{4l+6}
    \nabla^2\right)\bm\nabla^{(l)}\,\mf G(\mf r, \mf R).
\end{align}
Thus we see that the only terms surviving in the summation correspond to $l$ and the
$l-2$ with a Laplacian and all the other terms go to zero by biharmonicity or
due to odd powers of $\bm \rho$. Following the same method, it is
straightforward to show that boundary integral contribution of the stress tensor 
to the flow is
\begin{align}
    \bm K^{(l+1)}(\mf r,\,\mf R_m)
    =&
    \frac{4 \pi a^{l+1}\bm\Delta^{(l)} }{(l-1)!(2l +1)!!}
    \left( 1 + \frac{a^2}{4l+6}    \nabla_m^2\right)
    {\bm \nabla_m}^{(l-1)}
    \mf K(\mf r, \,\mf R_m),
\end{align}
%
%
%%
% ============================================================================================
\section{Irreducible parts of boundary integrals
}\label{appendix:solution_green_function}
% ============================================================================================
The flow due to the boundary integral of the Green's function is easily decomposed, using 
(\ref{eq:irred_traction_coefficients}), into its irreducible parts as
\begin{align}
   8 \pi \eta \,
  \mf v^{\mrm G}(\mf r)
  =& 
  -a^l\sum_{l=0}^{\infty}
    \left(1+\frac{a^{2}}{4l+6}\nabla^{2}\right)
  \bm\nabla^{(l)} \mf G 
  \odot 
  \mf{F}^{(l)},
  \nonumber\\
  =& 
  -a^l\sum_{l=0}^{\infty}
  \Bigg[
    \left(1+\frac{a^{2}}{4l+6}\nabla^{2}\right)
  \bm\Delta^{(l+1)}\bm\nabla^{(l)} \mf G 
  \odot 
  \mf{F}^{(l0)}
  \nonumber\\
 &-\frac{l}{l+1}
  \bm\Delta^{(l)}\left(\bm\nabla^{(l)} \times\mf G \right)
  \odot 
  \mf{F}^{(l1)}
  \nonumber\\
 &+\frac{l(l+1)}{2(2l+1)}
  \bm\Delta^{(l)}\bm\nabla^{(l-2)}\nabla^2 \mf G 
  \odot 
  \mf{F}^{(l2)}
    \Bigg],
\end{align}
The boundary integral of the stress tensor can be reduced to the same form 
by using (\ref{eq:irred_velocity_coefficients}).
Writing the double layer contribution in the index notation. 
\begin{align} 
   8 \pi \eta \,
  v^{\mrm K}_i(\mf r)&
  =   \eta  
  \sum_{l = 0}^{\infty}\left[
    \left(1+\frac{a^{2}}{4l+6}\nabla^{2}\right)
\nabla_{\alpha_1}\dots\nabla_{\alpha_{l-1}}
  K_{jik}
  V_{jk\alpha_1\dots\alpha_{l-1}} \right],
\end{align}
Now we use the (\ref{eq:irred_velocity_coefficients}) in the previous expression
to simplify it.
Lets first calculate the $\sigma=0$ contribution due to the double layer. Let consider
\begin{align}
\nabla_{\alpha_1}\dots\nabla_{\alpha_{l-1}}
  K_{jik}
    \Delta^{(l)}V_{jk\alpha_1\dots\alpha_{l-2}}^{(l0)} 
 &= 2\Delta^{(l)}
\nabla_{k}\nabla_{\alpha_1}\dots\nabla_{\alpha_{l-1}}G_{ij}
    V_{jk\alpha_1\dots\alpha_{l-2}}^{(l0)} ,
\label{}
\\
    \nabla_{\alpha_1}\dots\nabla_{\alpha_{l-1}}
    K_{jik}
    \Delta^{(l-1)}\epsilon_{jk\gamma} 
    V_{\gamma\alpha_2\dots\alpha_{l-1}}^{(l1)}
 &=\frac{2l-2}{l}
    \Delta^{(l-1)}\epsilon_{jk\gamma}\nabla_{k}\nabla_{\alpha_1}\dots\nabla_{\alpha_{l-1}}G_{ij},
    V_{jk\alpha_1\dots\alpha_{l-2}}^{(l1)} ,
    \label{}
\\
    \nabla_{\alpha_1}\dots\nabla_{\alpha_{l-1}}
    K_{jik} \Delta^{(l-1)}\delta_{jk}
    V_{\alpha_2\dots\alpha_{l-2}}^{(l2)}
    &= 
    -3 \Delta^{(l-1)}\nabla_{\alpha_3}\dots\nabla_{\alpha_{l-1}}
    \nabla^2G_{i\alpha_2}\label{}
    V_{\alpha_2\dots\alpha_{l-2}}^{(l2)}.
\end{align}
So all the three terms contribute from all the $three$ of its parts.
This can then be generalized to any $l\sigma$ as 
\begin{align}
   8 \pi \eta \,
  \mf v^{\mrm K}(\mf r)
  =&\frac{4\pi a^{(l+1)}}{(l-1)!(2l+1)!!} 
  \sum_{l=0}^{\infty}
  \Bigg[
    \left(1+\frac{a^{2}}{4l+6}\nabla^{2}\right)
  \bm\Delta^{(l+1)}\bm\nabla^{(l)} \mf G 
  \odot 
  \mf{V}^{(l0)}
  \nonumber\\
 &-\frac{l(2l-2)}{l(l+1)}
  \bm\Delta^{(l)}\left(\bm\nabla^{(l)} \times\mf G \right)
  \odot 
  \mf{V}^{(l1)}
  \nonumber\\
 &-\frac{3l(l+1)}{2(2l+1)}
  \bm\Delta^{(l)}\bm\nabla^{(l-2)}\nabla^2 \mf G 
  \odot 
  \mf{V}^{(l2)}
    \Bigg],
\end{align}
The two expressions due to single layer and double layer can then be added to
obtain the fluid flow (\ref{eq:galerkin_representation_flow_yl}) in terms of the
$\mf Q^{(l\sigma)}$. The expressions of $\mf Q^{l\sigma}$ is given in
(\ref{eq:generalized_Qlsigma}). We also provide the integral expressions below
\begin{subequations} \label{eq:single_sphere_Laddyzhenskaya_relations}
  \begin{align}
  \label{eq:single_sphere_Laddyzhenskaya_Stresslet}
    a\mf{Q}^{(20)} 
    =&
    \int \left[
    \frac12\left(\mf{f}^{} \, {\bm{\rho}} 
    + \left( \mf{f}^{} \, {\bm{\rho}} \right)^{{T}}\right)
    -\eta\,\left(
    \mf{v}^{} \, \widehat{\bm{\rho}} 
    + \left( \mf{v}^{} \, \widehat{\bm{\rho}} \right)^{{T}}\right)
    \right]\mrm dS,
\\
  \label{eq:single_sphere_Laddyzhenskaya_septlet}
a^2\mf{Q}^{(30)}
    =&\int
    \left[\frac16
    \left(
    \overbrace{\, \mf{f}^{} \, {\bm{\rho}} {\bm{\rho}} \,}
    - \frac25 (\mf{f}^{}\cdot {\bm{\rho}}) 
    \overbrace{{\bm{\rho}} \, \mathbb{I} }
    - \frac15 \overbrace{\mf{f}^{}\, \mathbb{I} }\right)
    -\frac{2\eta a}{3}\left(
    \overbrace{\, \mf{v}^{} \, \widehat{\bm{\rho}} \widehat{\bm{\rho}} \,}
    - \frac25 (\mf{v}^{}\cdot \widehat{\bm{\rho}}) 
    \overbrace{\widehat{\bm{\rho}} \, \mathbb{I} }
    - \frac15 \overbrace{\mf{v}^{}\, \mathbb{I} }
    \right)
    \right]\mrm dS,
\\
  \label{eq:single_sphere_Laddyzhenskaya_vortlet}
a^2\mf Q^{(31)}
    =& \int\left[\frac16\left(
    ( \mf{f}^{} \times {\bm{\rho}} ) \, {\bm{\rho}}
    + \left\{ ( \mf{f}^{} \times {\bm{\rho}} ) \, {\bm{\rho}}
    \right\}^{{T}}\right)
    -\frac{4\eta a}{9}\left(( \mf{v}^{} \times \widehat{\bm{\rho}} ) \, \widehat{\bm{\rho}}
    + \left\{ ( \mf{v}^{} \times \widehat{\bm{\rho}} ) \, \widehat{\bm{\rho}} \right\}^{{T}}
    \right)
    \right]\mrm dS,
\\
  \label{eq:single_sphere_Laddyzhenskaya_potDipole}
a^2\mf{Q}^{(32)} 
    =& \int \left[ \frac16
    \left((\mf{f}^{} \cdot {\bm{\rho}}) \, {\bm{\rho}} - \frac13
    \mf{f}^{}\right)
    -\eta a\left( (\mf{v}^{} \cdot \widehat{\bm{\rho}}) \,
    \widehat{\bm{\rho}} - \frac13 \mf{v}^{}\right)
    \right] \mrm dS,
\\
  \label{eq:single_sphere_Laddyzhenskaya_spinlet}
a^3\mf Q^{(41)}
   =&\int
   \left[\frac{1}{45}
   \left(
   \overbrace{\, \big(\mf{f^{}} \times {\bm{\rho}} \big) \, {\bm{\rho}} \, {\bm{\rho}} \,}
   - \frac35
   \overbrace{\big(\mf{f^{}} \times {\bm{\rho}} \big) \, \mathbb{I} \, }
   \right)
   -\frac{\eta a^2}{20}\left(
   \overbrace{\, \big(\mf{v^{}} \times \widehat{\bm{\rho}} \big) \, \widehat{\bm{\rho}} \, \widehat{\bm{\rho}} \,}
   - \frac35
   \overbrace{\big(\mf{v^{}} \times \widehat{\bm{\rho}} \big) \, \mathbb{I} \, }
   \right)
   \right]\mrm dS.
 \end{align}
\end{subequations}
\begin{table}
\begin{center}
\begin{tabular}{|c|c|c|c|}
\hline
& Single layer & Double layer & Total contribution\\
\hline
$C_{20} $                      & $\frac{20}{3}\pi \eta a^3 $ 
& $\frac83 \pi \eta a^3  $  & $\frac{28}{3}\pi \eta a^3 $ \\
\hline
$C_{21} $  & $-\frac12$       & & $-\frac12$\\                                              
\hline
$C_{30}$                          & $\frac{7}{6}\pi\eta a^5 \,$  
&$\frac{4}{15}\pi \eta a^5 \, $ & $ \frac{43}{30}\pi\eta a^5 $, \\
\hline
$C_{31}$                      & $\frac{8}{9}\pi \eta a^5\,$ 
& $\frac{32}{3}\pi\eta a^5\,$ & $\frac{13}{9}\pi\eta a^5$
\\
\hline
$C_{32} $              & $\frac45\pi\eta a^5 $ 
&-$\frac85\pi\eta a^5$ & $ -\frac{4}{5}\pi\eta a^5$ \\
\hline
$C_{41} $                        &- $\frac{1}{10}\pi \eta a^7 \,$ 
&- $12\pi \eta a^7 $ &-$\frac{121}{10}\pi\eta a^7  $\\
\hline
\end{tabular}
\end{center}
\caption{The coefficients $C_{l\sigma}$.}
\label{tab:c-l-sigma}
\end{table}
% ============================================================================================
\section{Evaluation of matrix elements}\label{appendix:matrix_elements}
% ============================================================================================
The expression of the diagonal ($m=n$) matrix elements can be obtained by
expanding the pressure vector, Green's function and the stress tensor in Fourier
series, and then evaluating the integrals of the matrix elements. The respective 
Fourier transforms of the pressure vector, Green's function and the stress
tensor are
\begin{subequations}
\begin{align} 
\label{eq:pressure_fourier}
\mf p (\mf k) = \frac{-i\, 8\pi \unv k}{k},
\end{align}
\begin{align}
\label{eq:greens_function_fourier}
\mf G (\mf k) = 
8\pi\bigg[\frac{\mathbb I -\unv{k} \unv{k}}{k^2}\bigg],
\end{align}
\begin{align}
\label{eq:double_layer_fourier}
\mf K(\mf k) &= 
        i\, 8\pi\bigg[ \frac{ \overbrace{\unv k\mathbb I} - 
        2\unv{k}\unv{k}\unv{k}}{k} \bigg],
\end{align}
\end{subequations}
where $\overbrace{\dots} $ implies complete symmetrization and $i$ is the unit
imaginary number. The single layer diagonal elements can be then written as
\begin{align}
\bm{G}_{nn}^{(l+1,\,l'+1)}
&= 
    \frac{(2l+1)(2l'+1)}{(4\pi a^2)^2}\,  
    \int\frac{\mrm d\mf k\, e^{i(\bm\rho-\bm\rho')\cdot \mf k}}{(2\pi)^3} 
    {\mf Y}^{(l)} (\bm{\rho}) \, 
    8\pi\left(\frac{\mathbb I -\unv{k} \unv{k}}{k^2}\right)
    {\mf Y}^{(l')} (\bm{\rho'}) \, \mrm{d} S_n \mrm{d} S_n. 
\end{align}
The integration can be performed using the orthogonality of the tensorial spherical 
harmonics ({\ref{eq:tensorial_spherical_harmonics_orthogonality}) and by
doing the plane wave expansion
\begin{align} \label{eq:plane_wave_expansion}
  e^{i \mf{k} \cdot \bm{\rho}} 
  &= 
  \sum_{q = 0}^{\infty} 
  \frac{(i)^q (2q+1)}{q!\left(2q-1\right)!!} 
  j_q(k \rho) 
  \left(
  \mf Y^{(q)}(\unv{k})
  \odot 
  \mf Y^{(q)}(\unv{\bm\rho}).
  \right)
\end{align}
We can then use the orthogonality relation of the spherical Bessel function
\begin{align}
  \int_0^{\infty} 
  j_l(ka) j_{l'}(ka) \,\mrm d k
  &= \delta_{ll'} 
  \frac{\pi}{2a\, (2l+1) },
\end{align}
to obtain diagonal-matrix elements of the single layer
\begin{align}
\bm{G}_{nn}^{(l+1,\,l'+1)}&= \delta_{ll'}
\,\frac{(2l+1)}{(2\pi a)}\,
\int 
\mf Y^{(l)}(\unv{\bm\rho})
\left(\mathbb I - \unv{\bm\rho} \unv{\bm\rho}\right)
\mf Y^{(l)}(\unv{\bm\rho}) \, d\Omega.
\end{align}
The double layer diagonal matrix elements are
\begin{align}
\bm{K}_{nn}^{(l+1,\,l'+1)}
&= 
    \frac{i 8\pi\,(2l+1)}{(4\pi a^2)\,l'!(2l'-1)!!}\,  
    \int\frac{ e^{i(\bm\rho-\bm\rho')\cdot \mf k}}{(2\pi)^3} 
    {\mf Y}^{(l)} (\bm{\rho}) \, 
        \bigg[ \frac{ \overbrace{\unv k\mathbb I} - 
        2\unv{k}\unv{k}\unv{k}}{k} \bigg]\cdot\unv{\bm\rho'}\,
    {\mf Y}^{(l')} (\bm{\rho'}) \,\mrm d\mf k\, \mrm{d} S_n \mrm{d} S_n. 
\end{align}
We now expand the plane wave in spherical Bessel function
(\ref{eq:plane_wave_expansion}). The diagonal matrix elements of the double layer is then
\begin{align}
\bm K_{nn}^{(l+1,\, l'+1)} & = i^{q-p+1}\frac{(2l+1)\left(2q+1\right)(2p+1)}{4\pi^3
a^2 \,l!\,q!p!\left( (2p-1)!! \right)^2(2q-1)!!}
    \nonumber\\
    &\int \mrm dS_n
    \mf Y^{(l)}( {\bm\rho})
    \mf Y^{(q)}( {\bm\rho})
    \int \mrm  d k \, k\, j_q(k a)   j_p(k a) 
    \nonumber\\
    &\int \mrm  d\bm\Omega_{\mf k}\,
    \mf Y^{(q)}( {\mf k}_n)
        \big[ \overbrace{\mf{\unv k}\mathbb{I}} - 
        2\mf{\unv k} \mf{\unv k}   \mf{\unv k} \big]
    \mf Y^{(p)}( {\mf k}_n)
    \cdot\int 
   \unv{\bm\rho'} \,
   \mf Y^{(l')}(\unv{\bm\rho'} )\,
   \mf Y^{(p)}(\unv{\bm\rho'} )\mrm dS_n.
\label{}
\end{align}
At this point we can use the orthogonality of tensorial spherical harmonics and the
following identity of the spherical Bessel function
\begin{align}
  \int_0^{\infty} 
  \mrm dk\, k\, j_p(ka) j_{q}(ka) 
  &= \delta_{p+1,q}
  \frac{\pi}{4a^2 }.
\end{align}
Also, the integration over the $m$-th surface can be performed 
using (\ref{eq:mazur1982_trick}) along with orthogonality relation of 
(\ref{eq:tensorial_spherical_harmonics_orthogonality}). The diagonal
contribution to the double layer thus reduces to %
\begin{align}\label{eq:appendix_diagonal_double}
\bm K_{nn}^{(l+1,\, l'+1)} & = -\frac{(2l+1)}{l!\,(2l'-1)!!}
    \int \mrm  d\bm\Omega_{\mf k}\,
    \mf Y^{(l')}( {\mf k}_n)
        \left[\unv{k}\cdot\left( \overbrace{\mf{\unv k}\mathbb{I}} - 
        2\mf{\unv k} \mf{\unv k}   \mf{\unv k} \right)\right]
    \mf Y^{(l)}( {\mf k}_n).
\end{align}

Now, (\ref{eq:mazur1982_trick}) and the orthogonality of the tensorial
spherical harmonics (\ref{eq:tensorial_spherical_harmonics_orthogonality})
can be used to simplify (\ref{eq:appendix_diagonal_double}). We also note
that $\unv{\bm k}\cdot \left( \overbrace{\mf{\unv k}\mathbb{I}} - 2\mf{\unv k}
\mf{\unv k}   \mf{\unv k} \right) = \mathbb I$. And hence the diagonal contribution of
double layer is $-4\pi\mathbb I$ at any order
\begin{align}
\bm K_{nn}^{(l+1,\, l'+1)} & = -\delta_{ll'}\,4\pi\mathbb I\, \bm\Delta^{(l)}.
\end{align}
The unknown traction coefficients can then be written in terms of known velocity
coefficients using (\ref{eq:tensorial_spherical_harmonics_orthogonality} and 
\ref{eq:mazur1982_trick})}.
\begin{align}
 8\pi \eta\,(\mf V-\mf V^a)  &= \left[
\,\frac{1}{(2\pi a)}\,
\int d\Omega\,
\mf Y^{(0)}(\unv{\bm\rho})
\left(\mathbb I - \unv{\bm\rho} \unv{\bm\rho}\right)
\mf Y^{(0)}(\unv{\bm\rho})\right]
\cdot \mf F^e
 = 
 \frac{4}{3a}\mf F^e.
\label{}
\end{align}
\begin{align}
 8\pi \eta\,(\mf \Omega-\mf\Omega^a)  &= \left[
\,\frac{3}{(4\pi a^3)}\,
\int d\Omega\,
\mf Y^{(1)}(\unv{\bm\rho})
\mf Y^{(1)}(\unv{\bm\rho})\right]\cdot
 \mf T^e
 =
 \frac{1}{a^3}\mf T^e,
\label{}
\end{align}
\begin{align}
 8\pi \eta\,\bm s  &= \left[
\,\frac{3}{(2\pi a^3)}\,
\int d\Omega\,
\mf Y^{(1)}(\unv{\bm\rho})
\left(\mathbb I - \unv{\bm\rho} \unv{\bm\rho}\right)
\mf Y^{(1)}(\unv{\bm\rho})\right]\odot
 \mf S
 =-
 \frac{6}{5a^3}\mf S.
\label{}
\end{align}
The expressions can be similarly calculated for higher $l$.
The off-diagonal elements can be calculated exactly by following the steps given in the 
boundary integral calculations (Appendix~\ref{appendix:flow_elements}). 
%
%%------------------------------------------------
%%
%%
%\bibliographystyle{unsrt}
%\bibliography{main_ref}
%%
%%

%
%%
%%%%
%%%%%%
%%%%%%%%%%

\end{document}